\documentclass{llncs}
\usepackage{makeidx}  % allows for indexgeneration
\usepackage{mathptmx}
\usepackage{amsmath}
\usepackage{amssymb}
\usepackage{graphicx}
\usepackage{times}
\usepackage{comment}

\graphicspath{{figs/}}                                  %% package figures
\newenvironment{pf}{\begin{proof}}{\qed\end{proof}}
\newcommand{\R}{\ensuremath{\mathbb{R}}}

\def\withcomments{
   
   \def\aside##1{}
   }
\withcomments

\begin{document}
\title{COAST: A Convex Optimization Approach\\to Stress-Based Embedding}

\author{
Emden R.~Gansner \and Yifan Hu \and Shankar Krishnan
}

\institute{
    AT\&T Labs - Research, Florham Park, NJ%\\
    %{\tt \{erg, yifanhu, krishnas\}@research.att.com}
}

\maketitle
\begin{abstract}
Visualizing graphs using virtual physical models is 
probably the most heavily used technique for drawing graphs in 
practice. There are many algorithms that are efficient and produce
high-quality layouts. If one requires that the layout also respect
a given set of non-uniform edge lengths, however, force-based approaches
become problematic while energy-based layouts become intractable.
In this paper, we propose a reformulation of the stress function
into a two-part convex objective function to which we can apply
semi-definite programming (SDP).
We avoid the high computational cost associated with SDP
by a novel, compact re-parameterization of the objective function using the
eigenvectors of the graph Laplacian. This sparse representation makes our approach scalable. 
We provide experimental results to show that 
this method scales well and produces reasonable layouts while  
dealing  with the edge length constraints.
\end{abstract}

\section{Introduction\label{sec_intro}}

For visualizing general undirected graphs,  
algorithms based on virtual physical models 
are some of the most frequently used drawing methods.
Among these, the spring-electrical model \cite{Eades_1984,Fruchterman_Reingold_1991}
treats edges as springs that pull nodes together, and nodes as
electrically-charged entities that repel each other.
Efficient and effective implementations \cite{Hachul_Junger_2005,Hu_2005,Walshaw_2003}
usually utilize a multilevel approach and fast force approximation with
a suitable spatial data structure, and can scale to 
millions of vertices and edges while still producing high-quality layouts.

In certain instances, the graph may assign non-uniform lengths to its edges, 
and the layout problem will have the additional
constraint of trying to match these lengths.
A suitable formulation of the spring-electrical model that
works well when edges have predefined target lengths is
still an open problem.
\begin{comment}
While it is possible to encode edge lengths in the attractive and 
repulsive forces (e.g.,~\cite{Battista_etal_1999}, section 10.1), 
such treatment is often {\it ad hoc} at best.
\end{comment}

In contrast, the (full) stress model assumes that there are springs
connecting all vertex pairs of the graph. Assuming we have a graph
$G=(V, E)$, with $V$ the set of vertices and $E$ the set of edges,
the energy of this spring system is

\begin{equation}
\sum _{i,j\in V} w_{{ij}} \left(\left\|x_i - x_j\right\| - d_{{ij}}\right){}^2,\label{spring_model}
\end{equation}
where $d_{ij}$ is the ideal distance between
vertices $i$ and $j$, and $w_{{ij}}$ is a weight factor.
The weight factor can modify the impact of an error. Weights can be arbitrary but are
frequently taken as a negative power of $d_{ij}$, thus lessening the error for larger
ideal distances.
A layout that minimizes this stress energy is taken as an optimal layout 
of the graph. 
The justification for this is clear: in most cases,
it is not possible to find a drawing that respects all of the edge lengths,
while expression (\ref{spring_model}) is basically the weighted mean square error
of a drawing. (See also the work of Brandes and Pich \cite{Brandes_Pich_2008}.) 

The stress model has its roots in multidimensional scaling (MDS)
\cite{Kruskal_MDS_1964} which was eventually applied to
graph drawing \cite{Kamada_Kawai_1989,Kruskal_MDS_1980}. 
Note that typically we are given only the ideal distance between vertices
that share an edge, which is taken to be unit length for graphs
without predefined edge lengths. For other vertex pairs, a common practice is to 
define $d_{ij}$ as the length of a shortest path between vertex $i$ and $j$. 
Such a treatment, however, means that an all-pairs shortest path
problem must be solved. Johnson's algorithm \cite{Johnson_1977}
takes $O(|V|^2\log |V| + |V||E|)$ time, and $O(|V|^2)$ memory.
(A slightly faster, but still quadratic, algorithm is also known
\cite{Pettie_2004_ASAP}.)
For large graphs, such complexities make solving the full stress model
infeasible.

A number of techniques have been proposed for circumventing this problem,
typically focused on approximate solutions, using only a few computed
distances, or approximating the shortest path calculations.
Gansner et al.~\cite{Gansner_Hu_North_maxent_tvcg_2013} proposed another 
approach for solving the ``stress model'' efficiently, by
redefining the problem.
The key idea was to note that only the edge
distances are given, while using shortest path lengths for the remainder
is somewhat arbitrary, and could be replaced with some other constraint
that is faster to compute but still works in terms of layout quality.
This led them to propose a two-part modified stress function
\begin{equation}
\sum _{\{i,j\}\in E} w_{{ij}} \left(\left\|x_i - x_j\right\| - d_{{ij}}\right)^2 - \alpha H(x),\label{maxent}
\end{equation}
where the first term encodes the stress associated with the given distances,
and the second handles the remaining pairs.

In this paper, we also consider minimizing a two-part modified stress function. 
However, our formulation is such that the objective function is convex. 
More specifically, it is quartic in the positions of the nodes, and
can be expressed as a quadratic function of auxiliary variables, 
where each of the auxiliary variables is a product of positions. 
We solve the problem by projecting the
positions into a subspace spanned by the eigenvectors of the Laplacian, and 
transform the minimization problem into one of convex programming. 
We call our technique COAST (Convex Optimization Approach to STress-based 
emdedding).

The rest of the paper is organized as follows. In Section~\ref{sec_related}, 
we discuss related work.
Section~\ref{sec_alg} gives the COAST model, and discusses a way to 
solve the model by semi-definite programming. 
Section~\ref{sec_results} evaluates our algorithm experimentally by comparing 
it with some of the existing
fast approximate stress models.
Section~\ref{sec_conc} presents a summary and topics for further study.

\section{Related Work}\label{sec_related}

Most of the earlier approaches 
\cite{silva_2003_landmarkmds,Gansner_Koren_North_2005b,Brandes_Pich_2006,KhouryHKS_2012,Gansner_Hu_North_maxent_tvcg_2013}
for efficiently handling graph drawings with edge lengths 
relied on approximately minimizing the stress model,
typically using some sparse model \cite{Gansner_Koren_North_2005b}. 
One notable effort is that of PivotMDS of Brandes and Pich
\cite{Brandes_Pich_2006}. This is an approximation algorithm which
only requires distance calculations from all nodes to a few chosen
nodes.

While PivotMDS is very efficient and works well for some graphs, for
graphs such as trees, multiple nodes can share the same position.
Khoury et al. \cite{KhouryHKS_2012} approximate the solution of the
linear system in a stress majorization
procedure~\cite{Gansner_Koren_North_2005b} by a low-rank 
singular value decomposition (SVD). They
used a result of Drineas et al.~\cite{Drineas:2004:CLG} which states
that for a matrix with well-distributed SVD values, the SVD values and
left SVD vectors of the submatrix consisting of randomly sampled
columns of the original matrix are a good approximation to the
corresponding SVD values and vectors of the original matrix.  With
this result, they were able to calculate only the shortest paths from a
selected number of nodes, as in PivotMDS. The method avoided having nodes in
a tree-like graph being embedded into the same position by using the
exact (instead of approximate) right-hand-side of the stress
majorization procedure, using an observation that this right-hand-side
can be calculated efficiently for the special case of $w_{ij} =
1/d_{ij}$.

The work most akin to that presented here is the maxent-stress model~\cite{Gansner_Hu_North_maxent_tvcg_2013}.
That approach borrows from the principle of maximal entropy, which says that items should be placed uniformly
in the absence of constraints.
The model tries to minimize the local stresses, while
selecting a layout that maximizes the dispersion of nodes.
This leads to the function shown in expression~(\ref{maxent}),
where typically $H(x) = ln_{\{i,j\}\notin E}\|x_i-x_j\|$.
The authors introduce an algorithm, called force-augmented stress majorization,
to minimize this objective function.

Although it essentially ignores edge lengths, the
binary stress model of Koren and \c{C}ivril \cite{Koren_2008_binary_stress}
is stylistically related, in that the first term attempts to specify
edge lengths (as uniformly 0) and the second term has the effect of uniformly spacing
the nodes.
Specifically,
there is a distance of 0 between nodes sharing an edge, and a distance of 1 otherwise,
giving the model

\begin{equation*}
\sum_{\{i,j\} \in  E} \|x_i - x_j\|^2 +
    \alpha\sum_{\{i,j\} \notin  E} \left(\|x_i - x_j\| - 1\right)^2.
\end{equation*}

Similarly,
Noack \cite{Noack_2007_energy_model,Noack_2009} has proposed the LinLog model
and, more generally, the $r$-PolyLog model,

\begin{equation*}
\sum_{\{i,j\} \in  E} \|x_i - x_j\|^r - \sum_{i,j \in  V} ln \|x_i - x_j\|,
\end{equation*}

\noindent where, in particular, the second term is suggestive of the use of entropy
in the maxent-stress model.

The most significant attempt to use a force-directed approach for encoding edge 
distances was the GRIP algorithm~\cite{Gajer_Goodrich_Kobourov_2000}.
The multilevel coarsening uses maximal independent set based filtration, with
the length of an edge at a coarse level computed from lengths of its composite
edges.  On coarse levels, the algorithm uses a version of the Kamada-Kawai 
algorithm \cite{Kamada_Kawai_1989} applied to each node within a local 
neighborhood of the original graph, thus handling the relevant edge lengths. 
On the finest level, however, a localized Fruchterman-Reingold 
algorithm~\cite{Fruchterman_Reingold_1991} is used, with no modeling of
edge lengths.

In the area of data clustering, Chen and Buja~\cite{Chen_2009_LMDS} present LMDS,
a model based on localized versions of MDS. Algebraically, this reduces to
\begin{equation*}
\sum_{\{i,j\} \in  S} \left(\left\|x_i - x_j\| - d_{{ij}}\right.\right)^2 - t \sum_{(i,j) \notin  S} \|x_i - x_j\|,\label{lmds_energy}
\end{equation*}
where $S$ contains $\{i,j\}$ if node $j$ is among the $k$ nearest neighbors if $i$.
It is difficult to determine how scalable this approach is but some tests indicate it
is not appropriate for graph drawing.

\section{The COAST Algorithm}\label{sec_alg}

Let \(G = (V, E)\) denote an undirected graph, with the
node set (vertices) \(V\) and edge set \(E\). We use \(n = |V|\)
for the number of vertices in \(G\).
We assume that each edge $(i,j)$ has a desired length 
$d_{ij}$ with weight $w_{{ij}}$.
Typically, one sets $w_{ij} = 1/d_{{ij}}{}^2$, but our analysis does not require that assumption.
We wish to embed $G$ into $d$-dimensional Euclidean space.
Let $x_i$ represent the coordinates of vertex $i$ in $\R^d$,
and let $P$ be the $n \times d$ matrix whose rows are the $x_i$.
We define the Gram matrix $X = (x_{ij})$ where $x_{ij} = x_i \cdot x_j$, the matrix of
inner products. It is well known that $X$ is a positive semi-definite matrix.

We consider minimizing a two-part modified stress function:

\begin{equation}
T(P) = \sum_{\{i,j\} \in  E} (w_{{ij}}\|x_i - x_j\|^2 - w_{{ij}}d_{{ij}}^2)^2 - t\lambda\sum_{(i,j) \notin  E} \|x_i - x_j\|^2,\label{energy}
\end{equation}
where the first term attempts to assign edges their ideal edge lengths, and the second term separates unrelated nodes as
much as possible. The parameter $t$ can be used to balance the two terms, emphasizing either conformity to
the specified edge lengths (small $t$) or uniform placement (large $t$). Without loss of generality, we can assume
a zero mean for the $x_i$, i.e., $\sum_{i} x_i = 0$. We 
set $\lambda = |E| / \left(\left(\begin{array}{c}n \\ 2\end{array}\right) - |E| + 1\right)$
to balance the relative size of the two terms, as suggested by Chen and 
Buja~\cite{Chen_2009_LMDS}.
To minimize $T(P)$, let $T_1$ and $T_2$ be the first and second terms of $T$,
respectively, so that $T = T_1 - T_2$, and consider the first term. 
We have the following derivation:
\begin{eqnarray}
T_1 & = & \sum_{\{i,j\} \in  E} {\{ w_{ij}(x_{ii} - x_{ij} - x_{ji} + x_{jj}) - w_{ij}{d_{ij}}^2\}}^2 \nonumber \\
 & = & \sum_{\{i,j\} \in  E} {\{ w_{ij}Tr(E_{ij}X) - w_{ij}{d_{ij}}^2\}}^2.
\label{f0}
\end{eqnarray}
where $Tr()$ is the trace function and $E_{ij} = (e_{kl})$ is the $n \times n$ matrix with
\begin{equation*}
e_{kl} = \left\{
\begin{array}{ll}
 1,& \text{if\ }k = l = i \text{\ or\ } k = l = j \\
 -1,& \text{if\ }k = i \text{\ and\ } l = j \\
 -1,& \text{if\ }k = j \text{\ and\ } l = i \\
0, & \text{otherwise}
\end{array}
 \right.
\end{equation*}
Using standard properties of the trace, the expression (\ref{f0}) can be rewritten as
\begin{equation}
\sum_{\{i,j\} \in  E} {{w_{ij}}^2 \{ {vec(E_{ij})}^T\mathcal{X} - {d_{ij}}^2\}}^2,
\label{f1}
\end{equation}
where $\mathcal{X} = vec(X)$ and $vec()$ is the matrix vectorization operator.

Functions defined on nodes of a graph can be well approximated by the eigenvectors of the
graph Laplacian \cite{Chung96}, and the smoother the function is, fewer eigenvectors are
required to approximate it well.
It is reasonable to assume that the function that embeds the vertices in $\R^d$ is smooth over 
the graph. Therefore, the bottom $k$ eigenvectors of the graph's Laplacian provide a good sparse basis for the 
position vectors. Typical values of $k$ range from 10-30 depending on the size of the graph. Let $Q \in \R^{n \times k}$ 
be the matrix composed of the eigenvectors of the Laplacian corresponding to the $k$ smallest eigenvalues, ignoring the eigenvalue 0. It is well known that the eigenvector corresponding to 
eigenvalue 0 accounts for the center of mass of the function. 
Removing it from consideration automatically places the embedding at the origin.
We can then find $k$ vectors $y_l$ in $\R^k$ so that we can write each $x_i$ 
as $\sum_{l} q_{il}y_l$ where $q_i = (q_{i1},q_{i2},\ldots,q_{ik})$ is the $i$th row of $Q$. 
If we then define the $k \times k$ positive semi-definite matrix $Y = (y_{ij})$ where $y_{ij} = y_i \cdot y_j$,
we have
\begin{displaymath}
X = P P^T = Q Y Q^T.
\end{displaymath}
Using $\mathcal{X} = vec(X)$ and letting $\mathcal{Y} = vec(Y)$, 
we can rewrite the above as 
\begin{displaymath}
\mathcal{X} = (Q \otimes Q) \mathcal{Y}, 
\end{displaymath}
where $\otimes$ is the Kronecker product.
Using this in expression (\ref{f1}), we have
\begin{equation}
T_1 = \sum_{\{i,j\} \in  E} {{w_{ij}}^2 \{ {vec(E_{ij})}^T(Q \otimes Q) \mathcal{Y} - {d_{ij}}^2\}}^2.
\label{f4}
\end{equation}

Since $x_i - x_j = \sum_{l} (q_{il} - q_{jl}) y_l$, it is fairly straightforward to see
that the following holds:

\begin{displaymath}
{vec(E_{ij})}^T(Q \otimes Q) = (q_i - q_j) \otimes (q_i - q_j).
\end{displaymath}

Applying this to equation (\ref{f4}), we have
\begin{eqnarray*}
T_1 & = & \sum_{\{i,j\} \in  E} {{w_{ij}}^2 \{ (q_i - q_j) \otimes (q_i - q_j) \mathcal{Y} - {d_{ij}}^2\}}^2 \\
  & = & \sum_{\{i,j\} \in  E} {w_{ij}}^2 \mathcal{Y}^T [{((q_i - q_j) \otimes (q_i - q_j))}^T ((q_i - q_j) \otimes (q_j - q_i)) ]\mathcal{Y} - \\
  &   & 2 \sum_{\{i,j\} \in  E}  {w_{ij}}^2 {d_{ij}}^2 ((q_i - q_j) \otimes (q_i - q_j)) \mathcal{Y} + \sum_{\{i,j\} \in  E} {w_{ij}}^2 {d_{ij}}^4.
\end{eqnarray*}

Now, turning to the second term of $T(P)$, we have
\begin{eqnarray}
T_2 & = & t\lambda\sum_{(i,j) \notin  E} \|x_i - x_j\|^2 \nonumber \\
    & = & t\lambda\left\{ \sum_{i > j} \|x_i - x_j\|^2 - \sum_{\{i,j\} \in  E} \|x_i - x_j\|^2 \right\}.
\label{e3}
\end{eqnarray}

\begin{lemma}
$\sum_{i > j} \|x_i - x_j\|^2 = nTr(Y)$ and $\sum_{\{i,j\} \in  E} \|x_i - x_j\|^2 = ((q_i - q_j) \otimes (q_i - q_j))\mathcal{Y}.$
\label{lemma2}
\end{lemma}
\begin{pf}
Because the $x_i$ have zero mean, the first summation is equal to $n \sum_i \|x_i \|^2 = n Tr(X) = n Tr(Y)$.
\end{pf}

Using lemma \ref{lemma2}, we can rewrite equation (\ref{e3}) as
\begin{eqnarray*}
T_2 & = & t\lambda \{n Tr(Y) - ((q_i - q_j) \otimes (q_i - q_j))\mathcal{Y}\} \\
    & = &  t\lambda \{n {vec(I)}^T - \sum_{\{i,j\} \in  E}((q_i - q_j) \otimes (q_i - q_j)) \} \mathcal{Y}.
\end{eqnarray*}

Combining our recastings of the two terms of equation (\ref{energy}), we have:
\begin{eqnarray*}
T(P) & = & T_1 - T_2 \\
     & = & \mathcal{Y}^T \left[ \sum_{\{i,j\} \in  E} {w_{ij}}^2 \{ ((q_i - q_j) \otimes (q_i - q_j))^T ((q_i - q_j) \otimes (q_i - q_j)) \} \right]\mathcal{Y} - \\
     &   & \left[ \sum_{\{i,j\} \in  E} (2 {w_{ij}}^2 {d_{ij}}^2 - t\lambda) ((q_i - q_j) \otimes (q_i - q_j)) - n t \lambda {vec(I)}^T \right] \mathcal{Y} + \\
     &   & \sum_{\{i,j\} \in  E} {w_{ij}}^2 {d_{ij}}^4.
\end{eqnarray*}

To simplify the exposition, we can write $T(P)$ 
as $\mathcal{Y}^T \mathbf{A} \mathcal{Y} + \mathbf{b}^T \mathcal{Y} + \text{constant}$. 
Since $A$ and $Y$ are symmetric positive semi-definite matrices, this is a convex function inside the 
semi-definite cone. It can be solved easily by any off-the-shelf semi-definite program (SDP). 
SDP is usually inefficient, taking cubic time in the size of the variables and constraints. 

A key novelty in our approach is the use of the approximation using the graph Laplacian. 
Instead of minimizing with $n^2$ variables, our re-parameterization with $Y$ reduces the 
number of variables to $k^2$. This is usually constant for most graphs and hence makes 
our approach scalable. 

Because of the special structure of our problem, we can further improve the running time
by converting our quadratically-constrained SDP to a Semidefinite Quadratic Linear Program (SQLP) 
and use a specialized solver such as SDPT3~\cite{TTT03}. 
If $\mathbf{A}$ is a symmetric positive semi-definite matrix
of rank $r$, we can use Cholesky decomposition to write $\mathbf{A} = \mathbf{R}^T\mathbf{R}$, where $\mathbf{R}$ is a $r \times k^2$ matrix.
Let $\mathcal{Z} = \mathbf{R} \mathcal{Y}$. Then we can rewrite $T(P)$ as:
\begin{equation*}
\min_{\alpha, \mathcal{Z}, \mathcal{Y}} \alpha + \mathbf{b}^T \mathcal{Y}
\end{equation*}
where $\mathcal{Z} = \mathbf{R} \mathcal{Y}$ and $\mathcal{Z}^T\mathcal{Z} \leq \alpha$.

We now show that the above optimization problem can be re-written as a linear function with {\bf one} 
second order cone constraint and few linear constraints. Let $\mathcal{K}_d$ 
denote the second order cone constraint of dimension $d$:

\begin{equation*}
\mathcal{K}_d = \{ (\beta, \mathbf{x}) \in \mathbb{R}^d: \parallel x \parallel \leq \beta \}
\end{equation*}

Then, if $\mathcal{W} = \left( \frac{1+\alpha}{2}, \frac{1-\alpha}{2}, \mathcal{Z}^T \right)^T$, it is
easy to see that $\mathcal{W} \in \mathcal{K}_{r+2}$. 
Now, we define vectors 
\begin{eqnarray*}
\mathbf{e_{-}}&=(1, -1, 0, \ldots, 0)^T & \in \mathbb{R}^{r+2} \\
\mathbf{e_{+}}&=(1, 1, 0, \ldots, 0)^T  & \in   \mathbb{R}^{r+2}  \\
\mathcal{V}&=(\mathcal{W}^T, \mathcal{Y}^T)^T & \in \mathbb{R}^{r+2+k^2}  \\
\mathbf{p}&=(\mathbf{e_{-}}^T, \mathbf{b}^T)^T & \in \mathbb{R}^{r+2+k^2}  \\
\mathbf{q}&=(\mathbf{e_{-}}^T, \mathbf{0}^T)^T & \in \mathbb{R}^{r+2+k^2} \\
\end{eqnarray*}
and the 
matrix $\mathbf{S} = (\mathbf{0}_{r \times 2}, -I_{r \times r}, \mathbf{R}) \in \mathbb{R}^{r \times r+2+k^2}$.

Then, with some simple algebraic manipulations, our optimization problem becomes
\begin{equation*}
\min_{\mathcal{V}} \mathbf{p}^T \mathcal{V}\\
\end{equation*}
where $\mathbf{q}^T \mathcal{V} = 1$, $\mathbf{S}\mathcal{V} = \mathbf{0}$ and
$\mathcal{W} \in \mathcal{K}_{r+2}$.

\section{Experimental Results\label{sec_results}}

We implemented the COAST algorithm in a combination of Python, Matlab and C code. 
The main parts consist of forming the matrix $A$ and vector $b$, 
calculating the eigenvectors of the Laplacian, and
solving the optimization problem. Time for the last part is dependent only on the 
number of eigenvectors $k$, hence is constant for a fixed number of eigenvectors.
For graphs of size up to $100,000$, the minimization 
using SQLP takes less than 10 seconds inside Matlab.

We tested the COAST algorithm for
solving the quartic stress model on a range of graphs.
For comparison, we also tested PivotMDS; PivotMDS(1), which uses PivotMDS,
followed by a sparse stress majorization;
the maxent-stress model Maxent; and the full stress model, using stress majorization. 
We summarize all the tested algorithms in Table~\ref{alg}.

    \begin{table} 
    \caption{\small\sf Algorithms tested.\label{alg}}

     \begin{center}
     \begin{tabular}{|c|c|c|c|c|c|c|c|c|c|c|c|c|c|c|c|c|c||c|c|c|c|c|c|}
\hline
 Algorithm        &  Model                                 & Fits distances? \\
\hline
 COAST             & quartic stress model                        & Yes. Edges only \\
\hline
 PivotMDS             & approx. strain model                   & Yes/No \\
\hline
 PivotMDS(1)          & PivotMDS $+$ sparse stress          & Yes.   \\
\hline
 Maxent        & PivotMDS $+$ maxent-stress                 & Yes.  \\
\hline	
 FSM              & full stress model                      & Yes. All-pairs  \\
\hline	
\end{tabular}
     \end{center}
     \end{table}

With the exception of graph {\tt gd}, which is
an author collaboration graph of the International Symposium on Graph Drawing between 1994-2007,
the graphs used are from the University of Florida Sparse
Matrix Collection~\cite{Davis_Hu_ufl_2009}.
Our selection is exactly the same as that used by Gansner et al.~\cite{Gansner_Hu_North_maxent_tvcg_2013}.
Two of the graphs ({\tt commanche} and {\tt luxembourg}) have associated
pre-defined non-unit edge lengths.
In our study, a rectangular matrix, or one with an asymmetric pattern,
is treated as a bipartite graph.
Test graph sizes are given in Table~\ref{graphs}.
     \begin{table} 
    \caption{\small\sf Test graphs. Graphs marked ${}^*$ have pre-specified non-unit edge lengths. Otherwise, unit edge length is assumed.\label{graphs}}
     \begin{center}
     \begin{tabular}{|c|r|r|c|c|c|c|c|c|c|c|c|c|c|c|c|c|c||c|c|c|c|c|c|}
     \hline   
 \text{Graph} & \multicolumn{1}{c}{$|V|$} & \multicolumn{1}{c}{$|E|$} & description \\
\hline
 \text{{\tt gd}} & 464 & 1311  &Collaboration graph\\
\hline
 \text{{\tt btree}} & 1023 & 1022  &Binary tree\\
\hline
 \text{{\tt 1138\_bus}} & 1138 & 1358  &Power system\\
\hline				   		 
 \text{{\tt qh882}} & 1764 & 3354  &Quebec hydro power\\
\hline				   		 
 \text{{\tt lp\_ship04l}} & 2526 & 6380  &Linear programming\\
\hline
 \text{{\tt USpowerGrid}} & 4941 & 6594 & US power grid\\
\hline				   		 
 \text{{\tt commanche}}${}^*$ & 7920 & 11880  & Helicopter\\
\hline				   		 
\text{{\tt bcsstk31}} & 35586 & 572913  & Automobile component\\
\hline				   		 
 \text{{\tt luxembourg}}${}^*$ & 114599 & 119666  & Luxembourg street map\\
\hline	
\end{tabular}
     \end{center}
     \end{table}

Tables~\ref{btree_key} -- \ref{heli_key} present the outcomes for all of the graphs.
Following Brandes and Pich \cite{Brandes_Pich_2008}, each drawing has an associated error chart.
In an error chart, the $x$-axis gives the graph distance bins, the
$y$-axis is the difference between the actual geometric distance in the layout and the graph
distance. The chart shows the median (black line), the 25 and 75 percentiles
(gray band) and the min/max errors (gray lines) that fall within each bin. 
For ease of understanding, we plot graph distance against
distance error, instead of graph distance vs. actual distance as
suggested by Brandes and Pich \cite{Brandes_Pich_2008}. 
Because generating the error chart requires an all-pairs
shortest paths calculation, we provide this chart only
for graphs with less than 10,000 nodes.

\begin{table}
\centering
\caption{\small\sf Drawings and error charts of the tested algorithms for  {\tt gd} and {\tt btree}.}
\label{btree_key} 
\begin{tabular}{|c|c|c|c|c|c|c|c|}
\hline   
\text{Graph} &\text{PivotMDS} & \text{PivotMDS(1)} & \text{Maxent} & \text{COAST}&\text{FSM}\\
\hline
 \raisebox{-0.75ex}[0pt]{\text{{\tt gd}}} & \includegraphics[width=2.3cm]{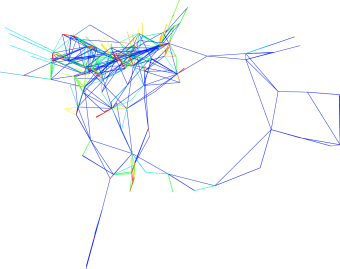} &\includegraphics[width=2.3cm]{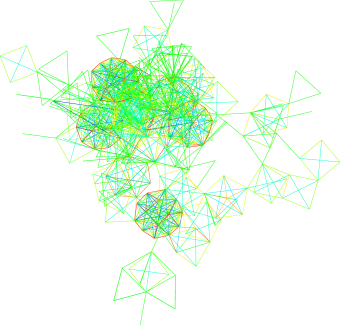}&\includegraphics[width=2.3cm]{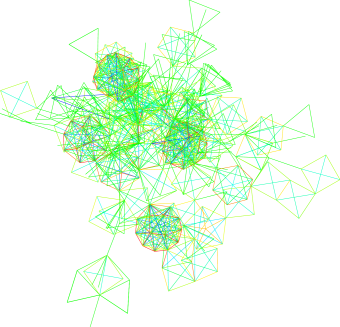}&\includegraphics[width=2.3cm]{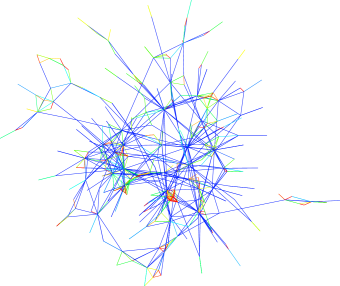} & \includegraphics[width=2.3cm]{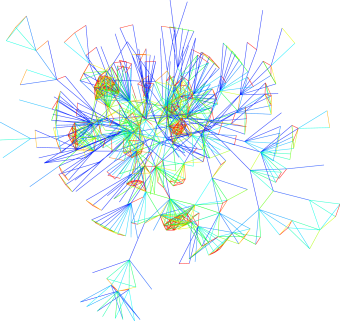}\\
\cline{2-6}
 & \includegraphics[width=2.3cm]{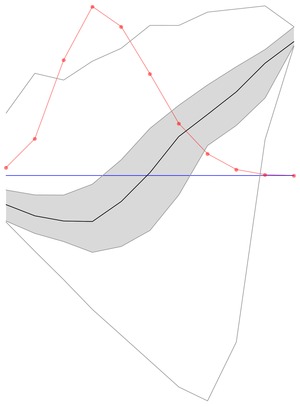} &\includegraphics[width=2.3cm]{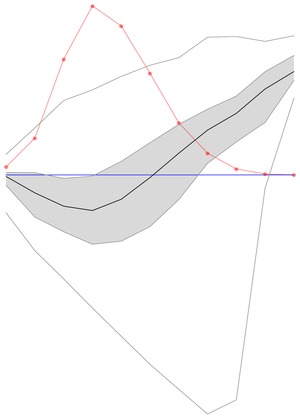}&\includegraphics[width=2.3cm]{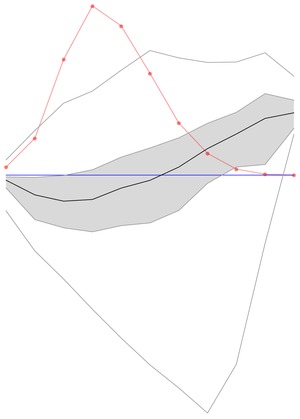} & \includegraphics[width=2.3cm]{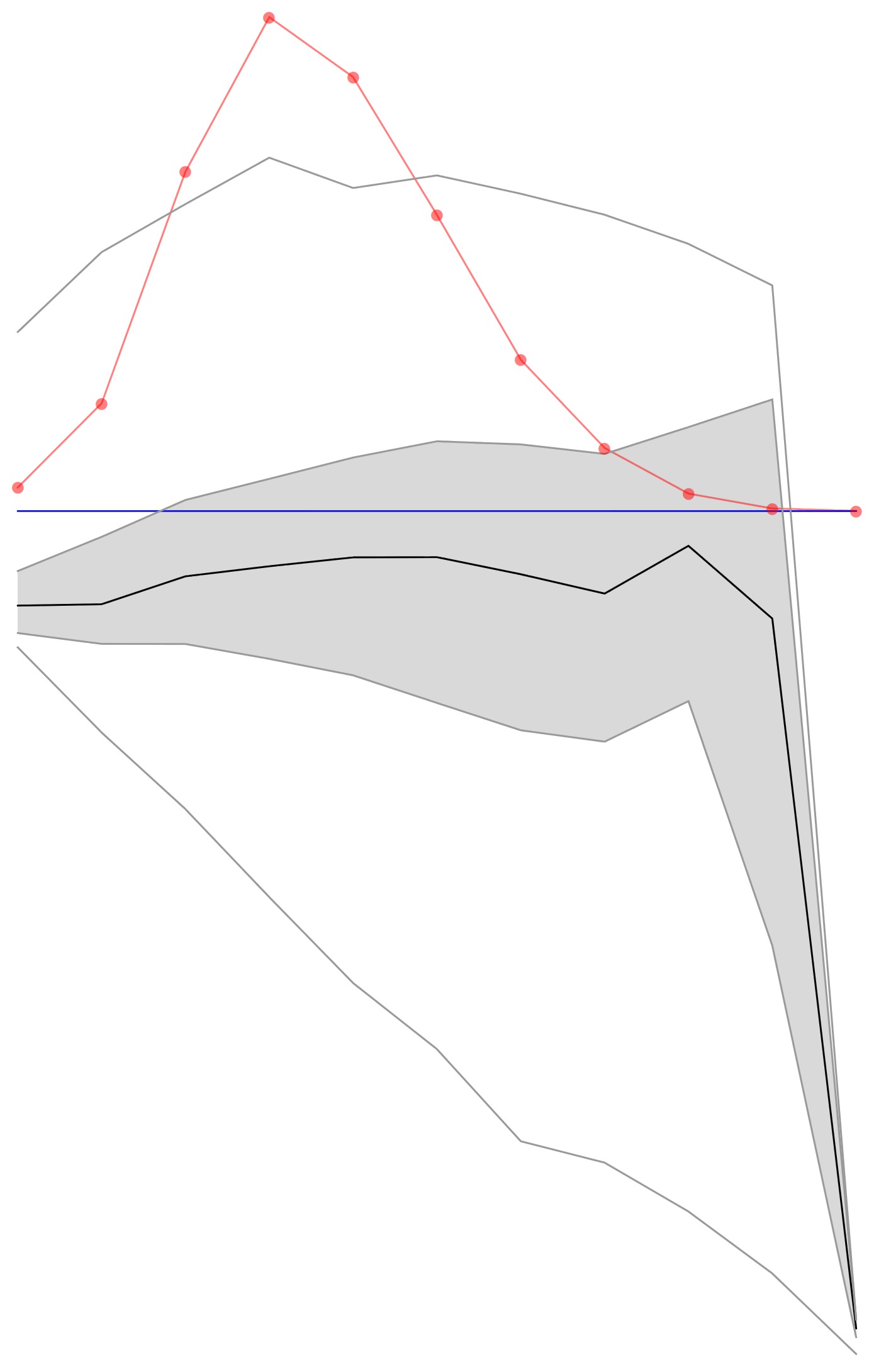}& \includegraphics[width=2.3cm]{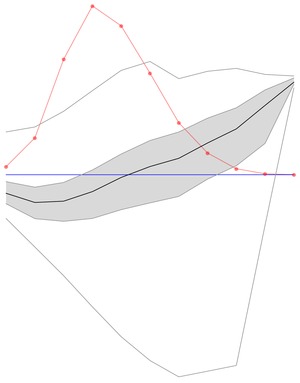}\\
\hline
 \raisebox{-0.75ex}[0pt]{\text{{\tt btree}}} & \includegraphics[width=2.3cm]{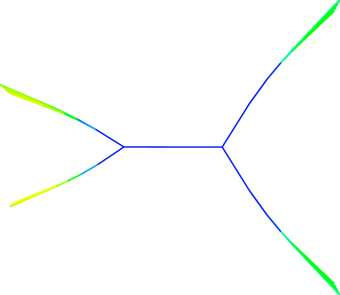} &\includegraphics[width=2.3cm]{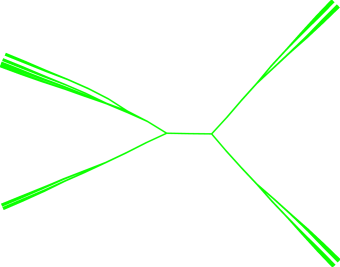}&\includegraphics[width=2.3cm]{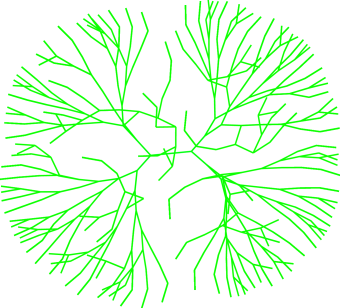} & \includegraphics[width=2.3cm]{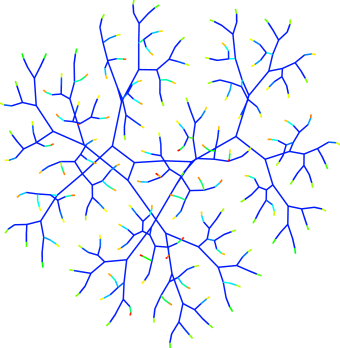} & \includegraphics[width=2.3cm]{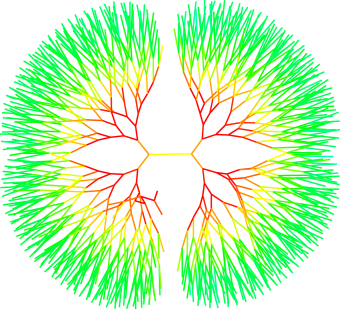}\\
\cline{2-6}
 & \includegraphics[width=2.3cm]{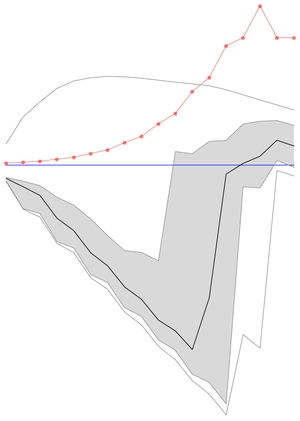} &\includegraphics[width=2.3cm]{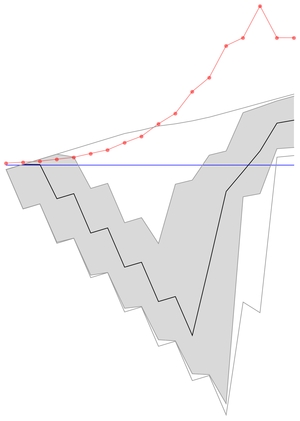}&\includegraphics[width=2.3cm]{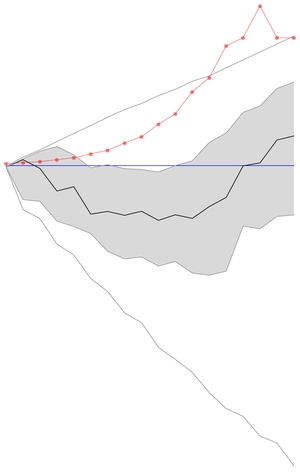}& \includegraphics[width=2.3cm]{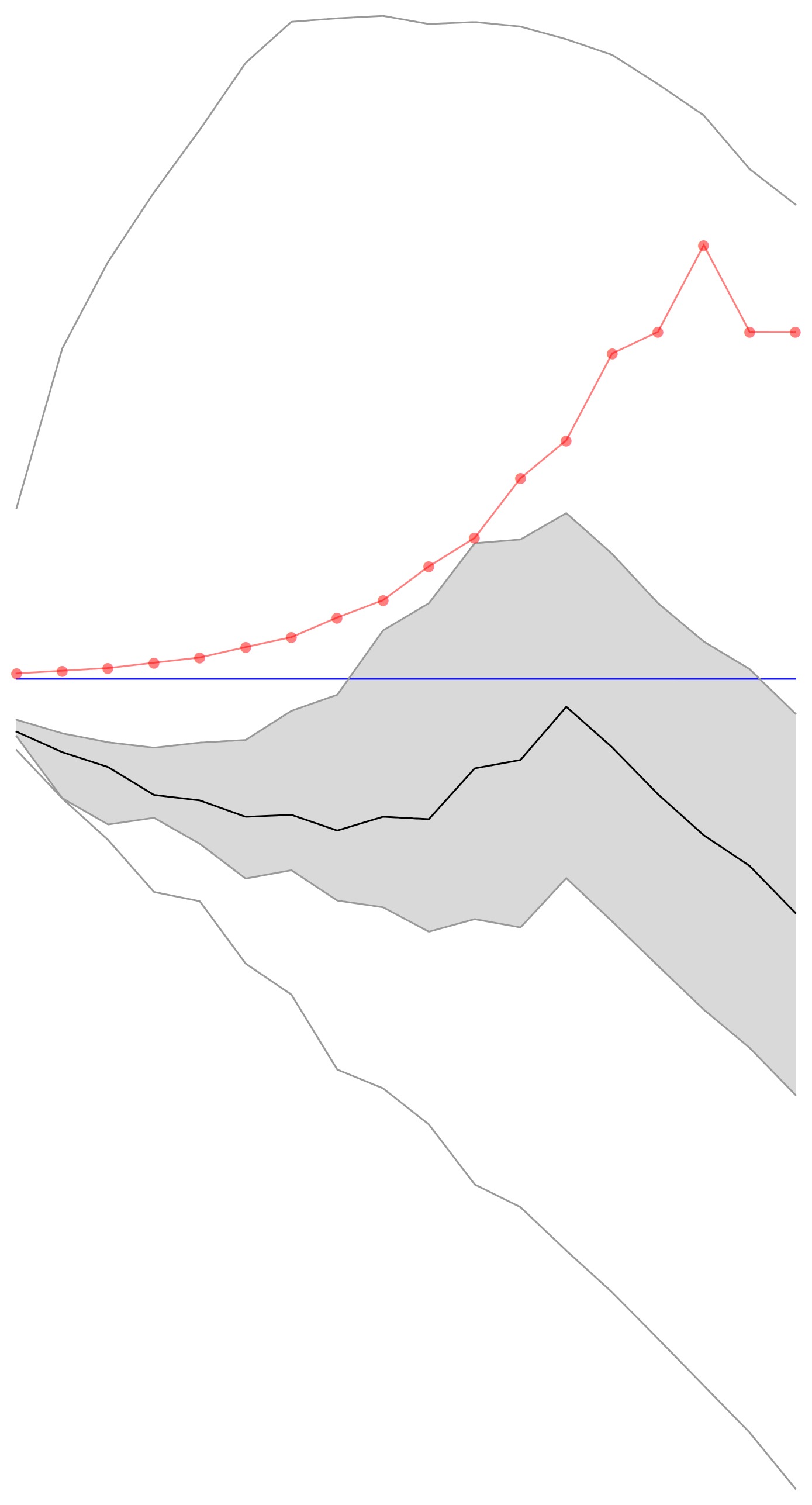}& \includegraphics[width=2.3cm]{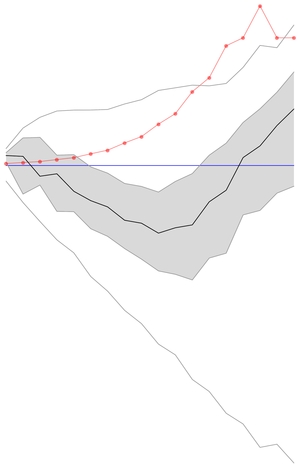}\\
\hline				   		 
\end{tabular}
\end{table}

\begin{table}
\centering
\caption{\small\sf Drawings and error charts of the tested algorithms for  {\tt 1138\_bus} and {\tt qh882}.}
\label{1138_key} 
\begin{tabular}{|c|c|c|c|c|c|c|c|}
\hline   
 \text{Graph} &\text{PivotMDS} & \text{PivotMDS(1)} & \text{Maxent} & \text{COAST}&\text{FSM}\\
\hline
 \raisebox{-0.75ex}[0pt]{\text{{\tt 1138\_bus}}} & \includegraphics[width=2.3cm]{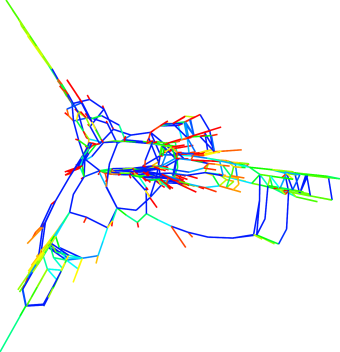} &\includegraphics[width=2.3cm]{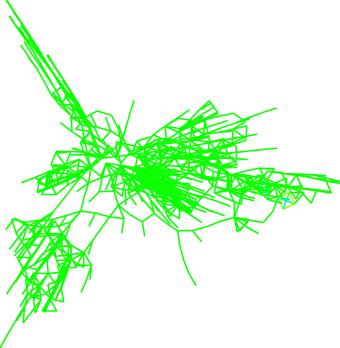}&\includegraphics[width=2.3cm]{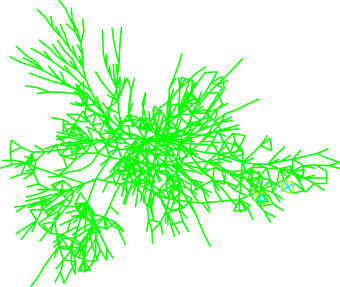} & \includegraphics[width=2.3cm]{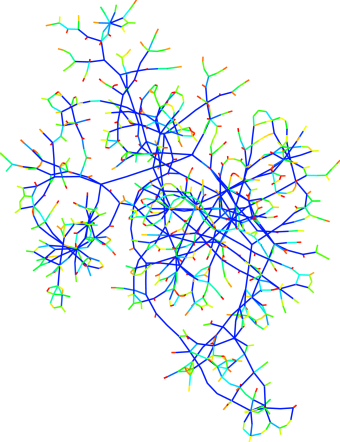} & \includegraphics[width=2.3cm]{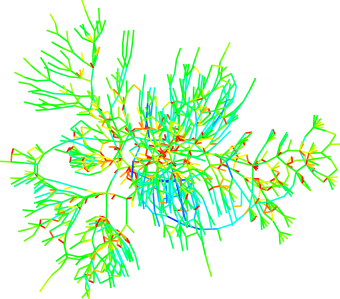}\\
\cline{2-6}
 & \includegraphics[width=2.3cm]{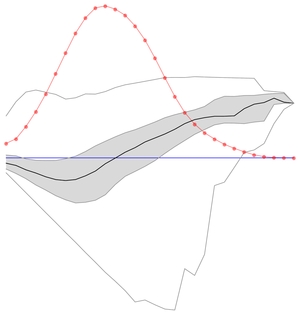} &\includegraphics[width=2.3cm]{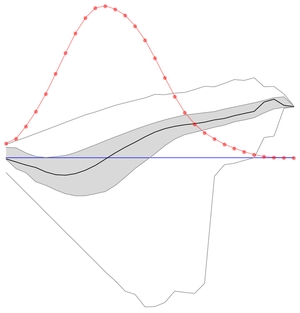}&\includegraphics[width=2.3cm]{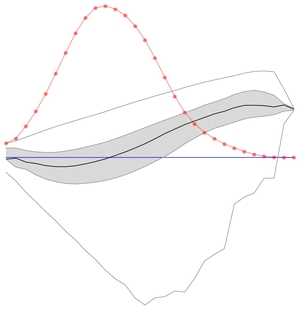} & \includegraphics[width=2.3cm]{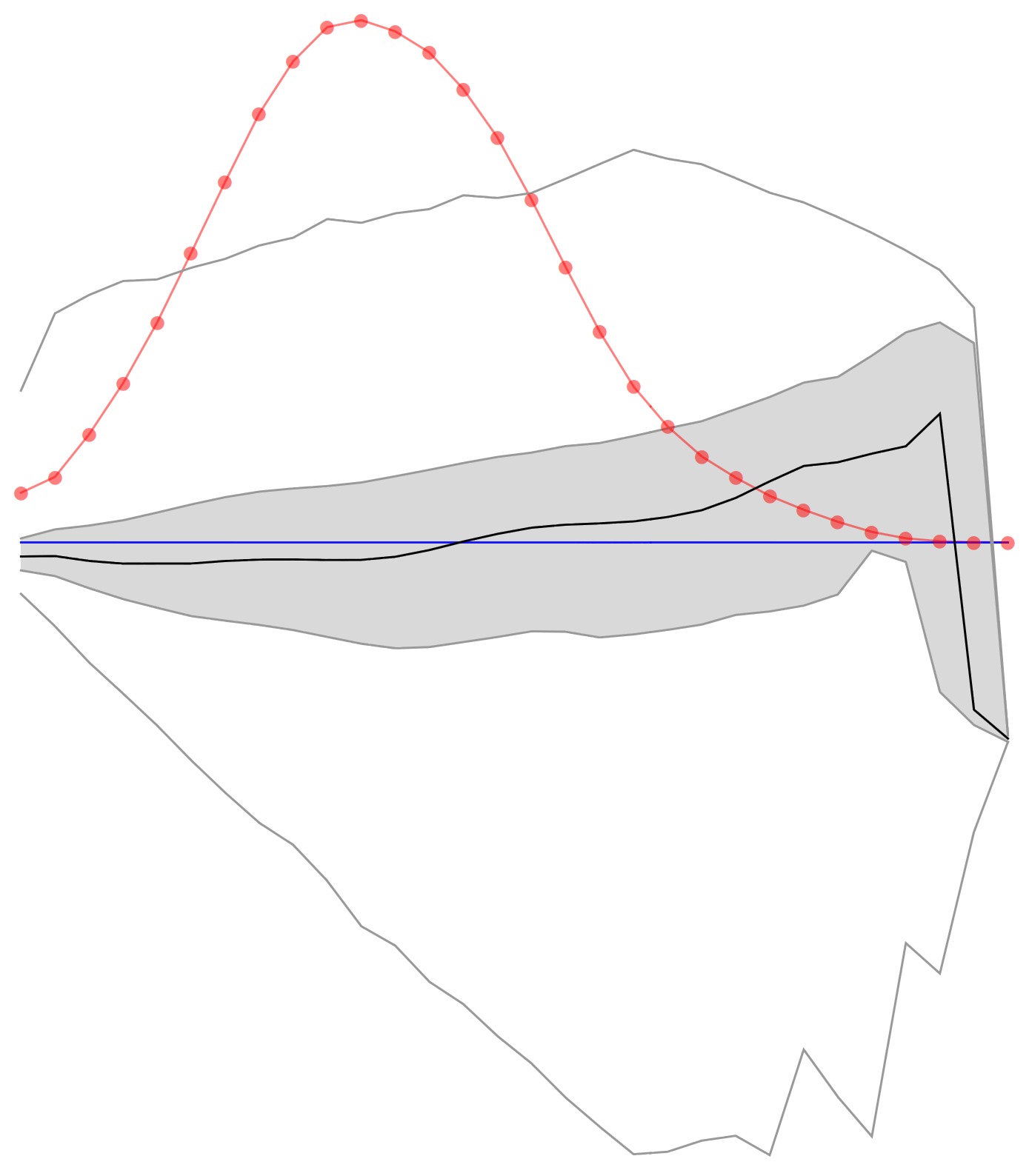}& \includegraphics[width=2.3cm]{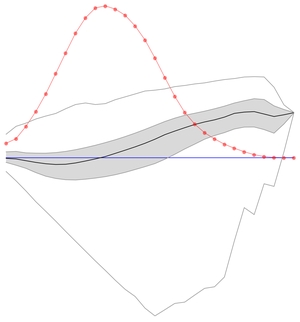}\\
\hline				   		 
 \raisebox{-0.75ex}[0pt]{\text{{\tt qh882}}} & \includegraphics[width=2.3cm]{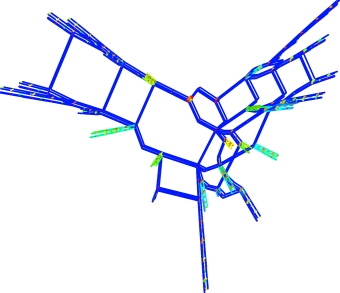} &\includegraphics[width=2.3cm]{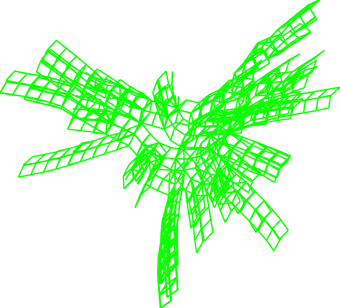}&\includegraphics[width=2.3cm]{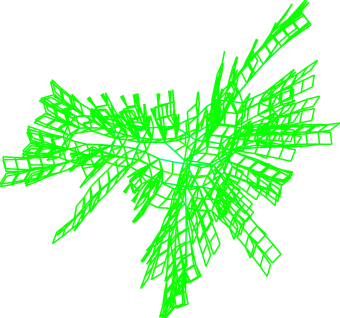} & \includegraphics[width=2.3cm]{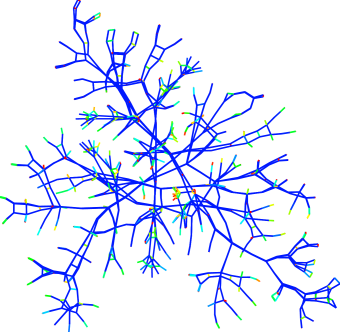}& \includegraphics[width=2.3cm]{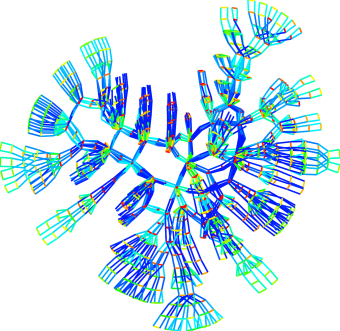}\\
\cline{2-6}				   		 
 & \includegraphics[width=2.3cm]{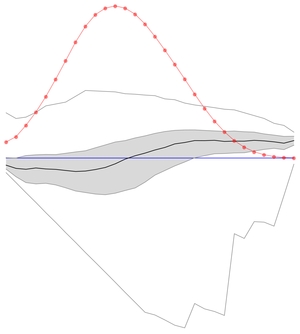} &\includegraphics[width=2.3cm]{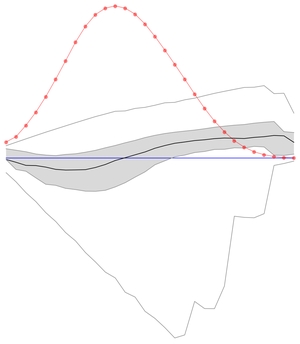}&\includegraphics[width=2.3cm]{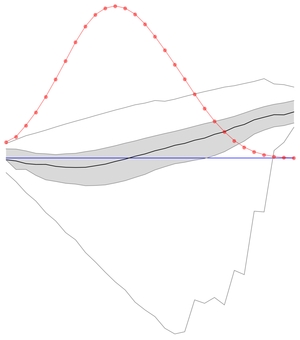} & \includegraphics[width=2.3cm]{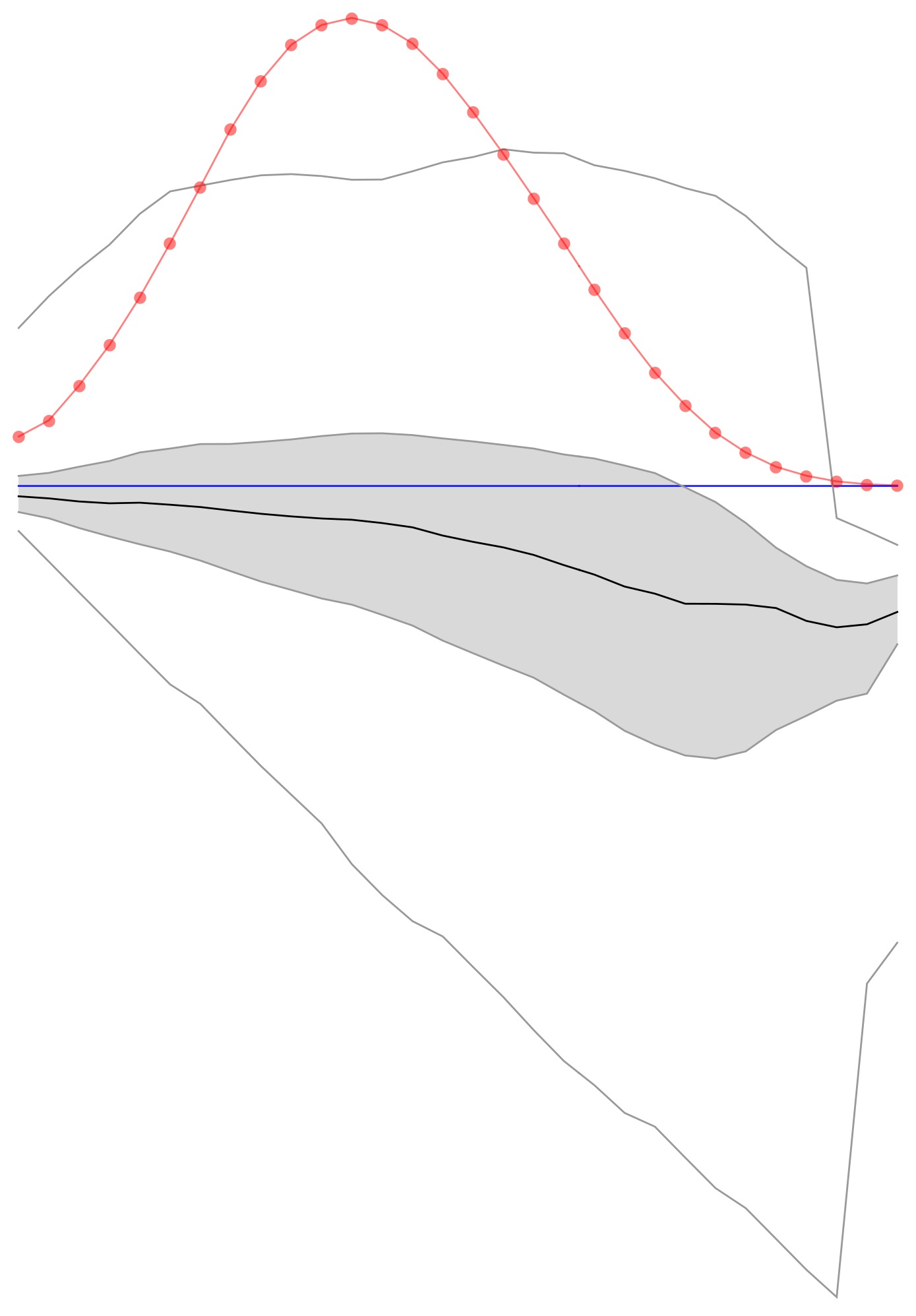}& \includegraphics[width=2.3cm]{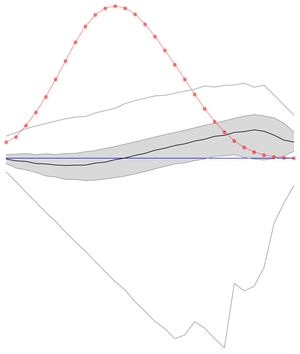}\\
\hline				   		 
\end{tabular}
\end{table}

\begin{table}
\caption{\small\sf Drawings and error charts of the tested algorithms for  {\tt lp\_ship041} and {\tt USPowerGrid}.}
\centering
\label{lpship_key} 
\begin{tabular}{|c|c|c|c|c|c|c|c|}
\hline   
 \text{Graph} &\text{PivotMDS} & \text{PivotMDS(1)} & \text{Maxent} & \text{COAST}&\text{FSM}\\
\hline				   		 			   		 
 \raisebox{-0.75ex}[0pt]{\text{{\tt lp\_ship04l}}} & \includegraphics[width=2.3cm]{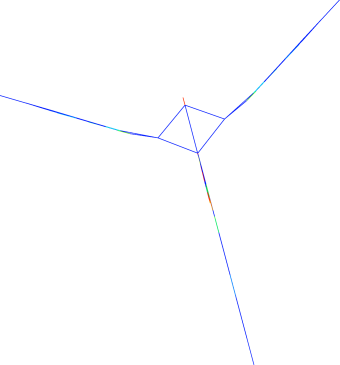} &\includegraphics[width=2.3cm]{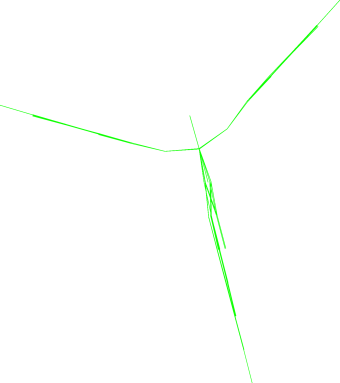}&\includegraphics[width=2.3cm]{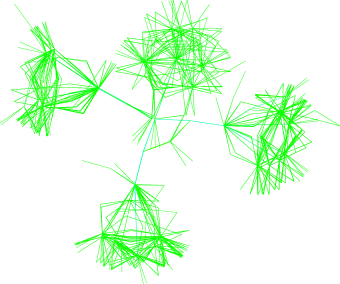}& \includegraphics[width=2.3cm]{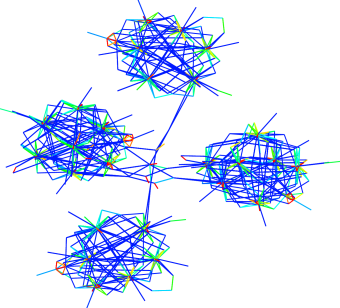} & \includegraphics[width=2.3cm]{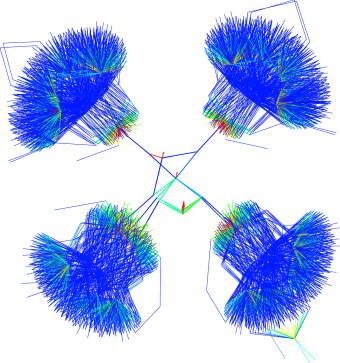}\\
\cline{2-6}				   		 
 & \includegraphics[width=2.3cm]{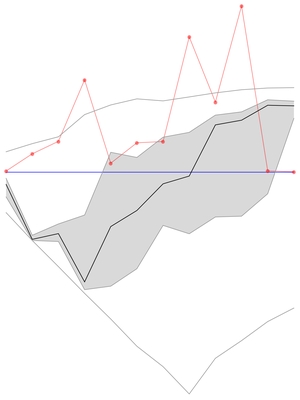} &\includegraphics[width=2.3cm]{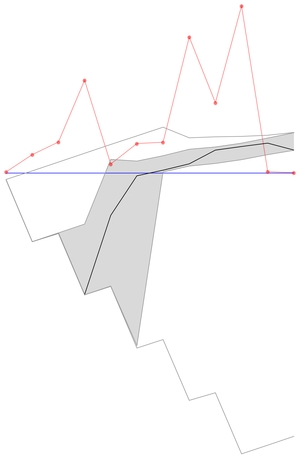}&\includegraphics[width=2.3cm]{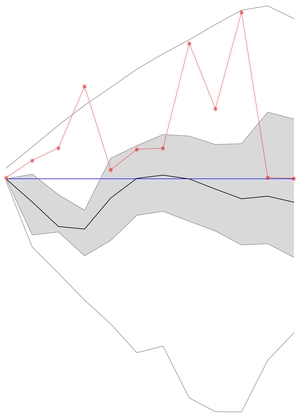} & \includegraphics[width=2.3cm]{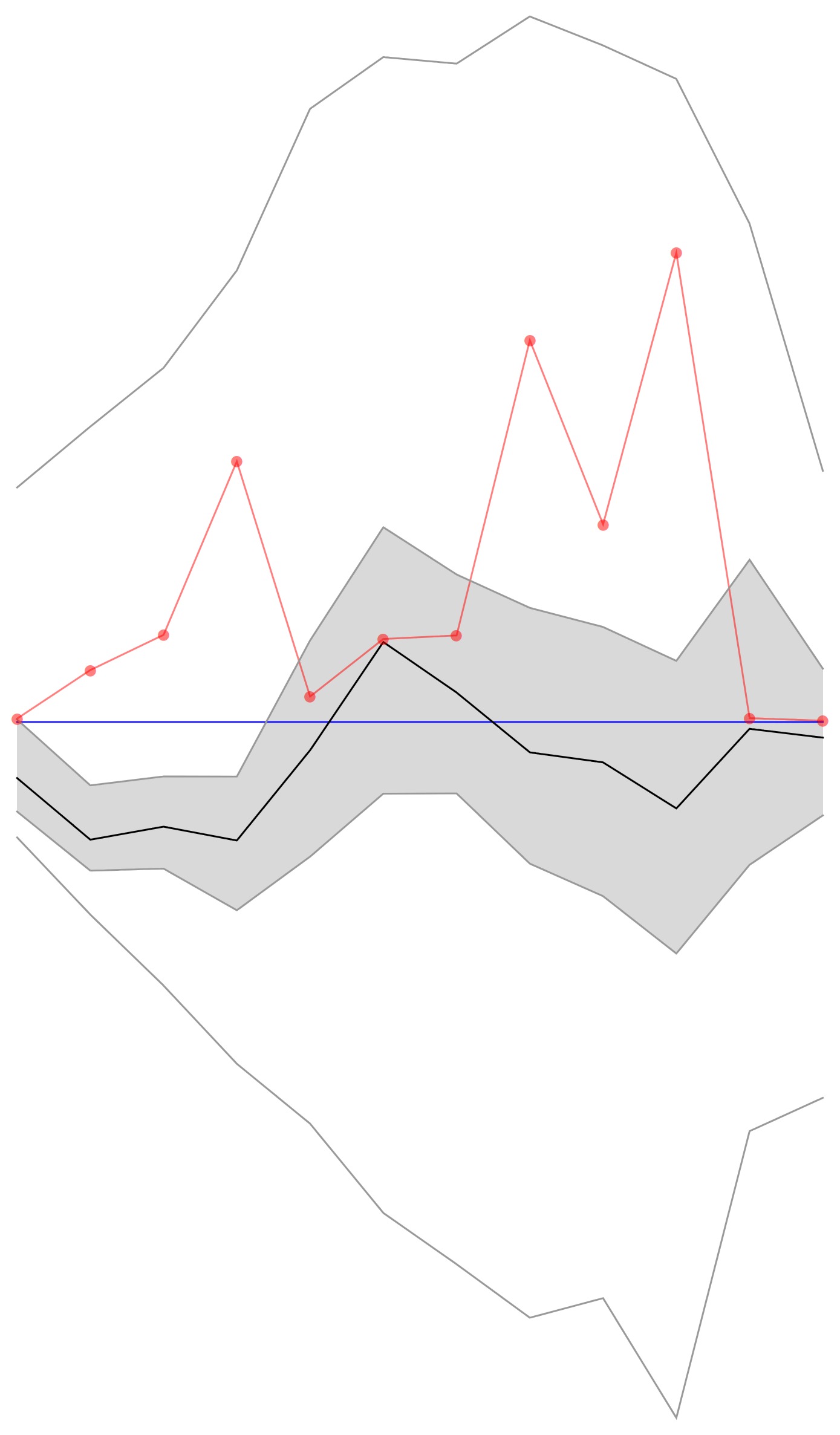}& \includegraphics[width=2.3cm]{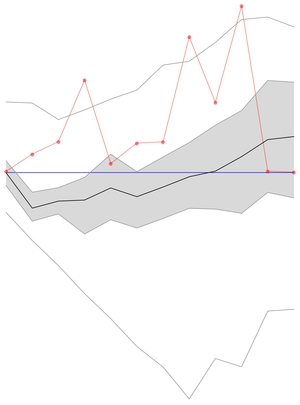}\\
\hline				   		 
  \raisebox{-0.75ex}[0pt]{\text{{\tt USpowerGrid}}} & \includegraphics[width=2.3cm]{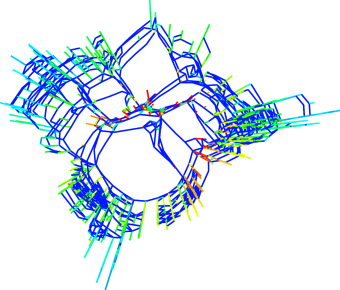} &\includegraphics[width=2.3cm]{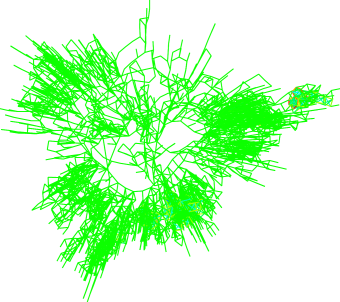}&\includegraphics[width=2.3cm]{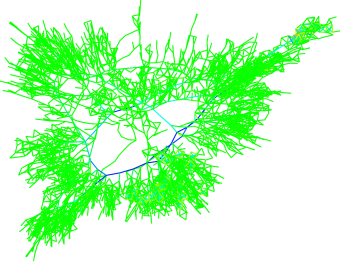} & \includegraphics[width=2.3cm]{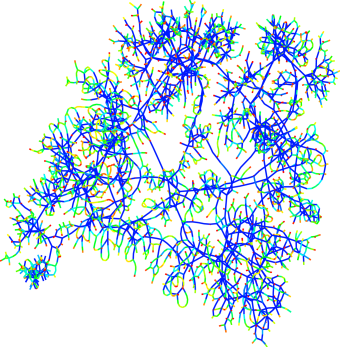} & \includegraphics[width=2.3cm]{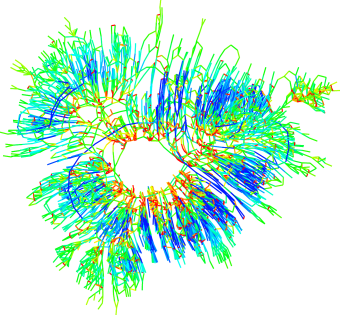}\\
\cline{2-6}				   		 
 & \includegraphics[width=2.3cm]{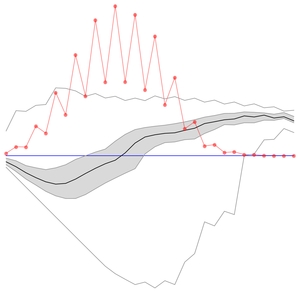} &\includegraphics[width=2.3cm]{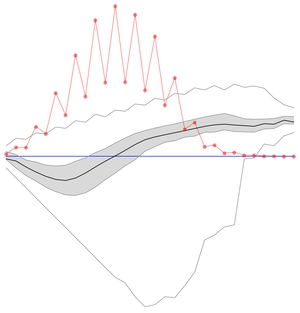}&\includegraphics[width=2.3cm]{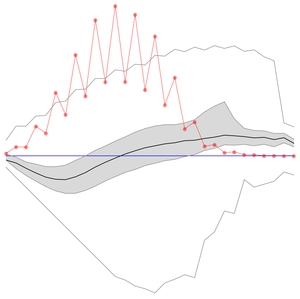} & \includegraphics[width=2.3cm]{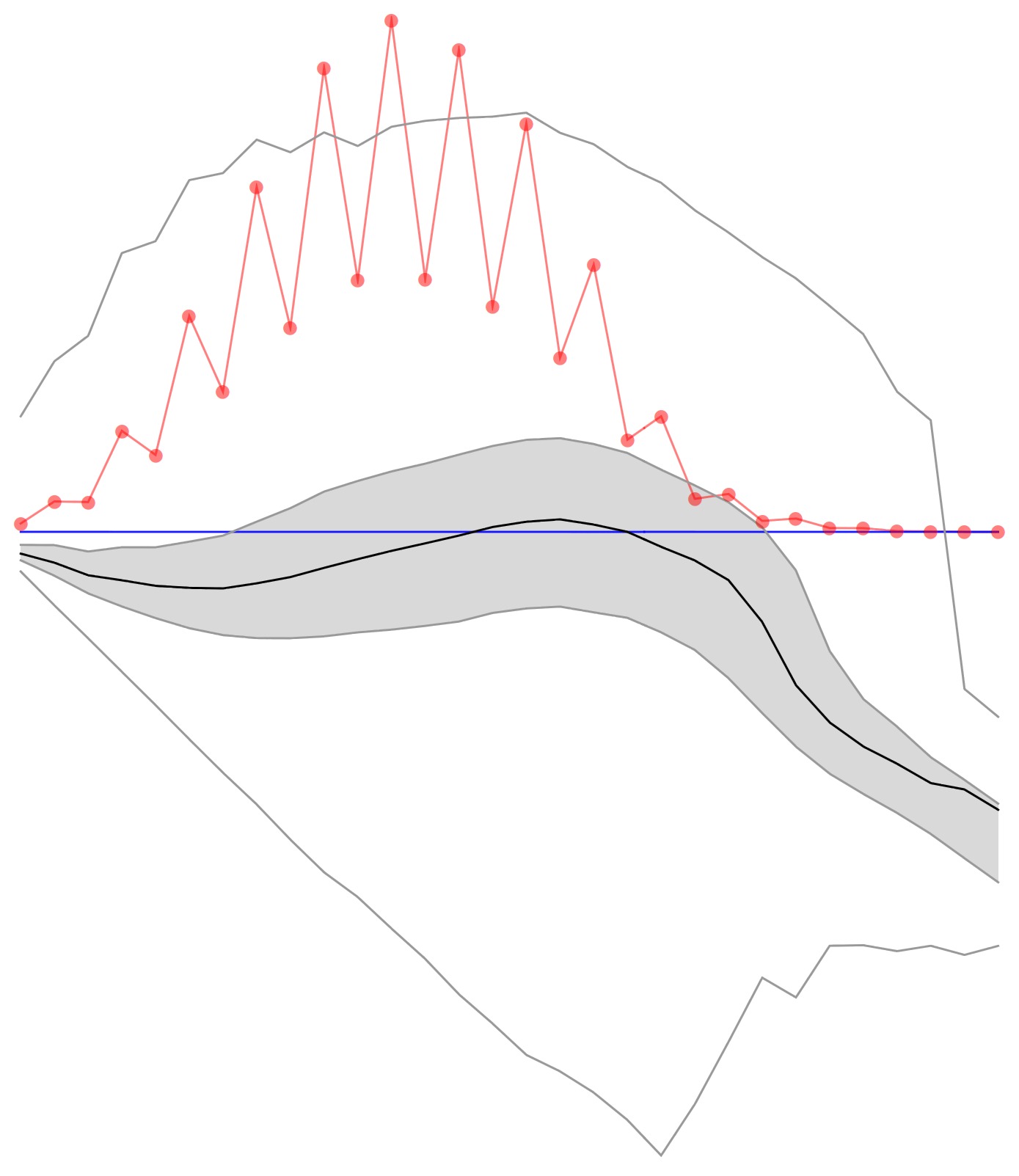}& \includegraphics[width=2.3cm]{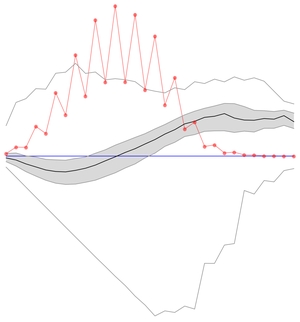}\\
\hline	
\end{tabular}
\end{table}

\begin{table}
\caption{\small\sf Drawings and error charts of the tested algorithms for  {\tt commanche}, {\tt bcsstk31} and {\tt Luxembourg}.}
\centering
\label{heli_key}
\begin{tabular}{|c|c|c|c|c|c|c|c|c|c|c|c|c|c|c|c|c|c||c|c|c|c|c|c|}
\hline
 \text{Graph} &\text{PivotMDS} & \text{PivotMDS(1)} & \text{Maxent} & \text{COAST}&\text{FSM}\\
\hline
 \raisebox{-0.75ex}[0pt]{\text{{\tt commanche}}} & \includegraphics[width=2.3cm]{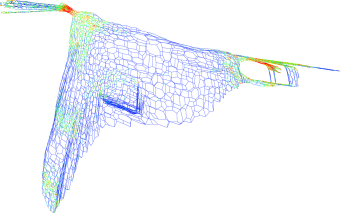} &\includegraphics[width=2.3cm]{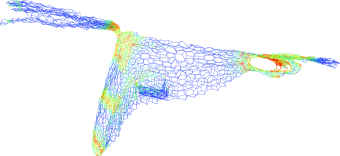}&\includegraphics[width=2.3cm]{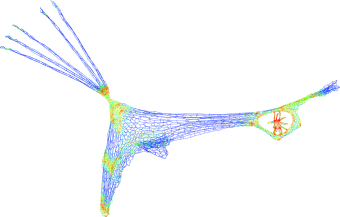} & \includegraphics[width=2.3cm]{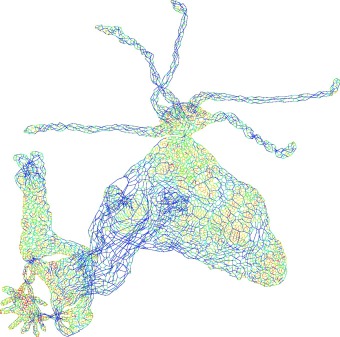} & \includegraphics[width=2.3cm]{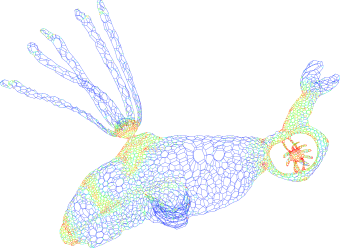}\\
\cline{2-6}				   		 
 & \includegraphics[width=2.3cm]{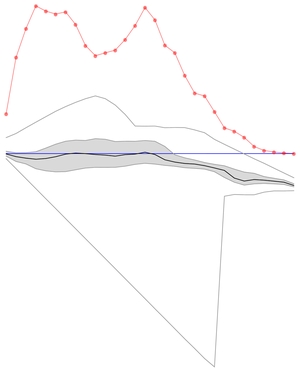} &\includegraphics[width=2.3cm]{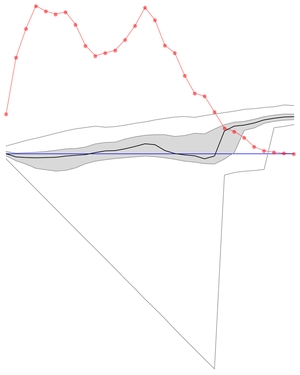}&\includegraphics[width=2.3cm]{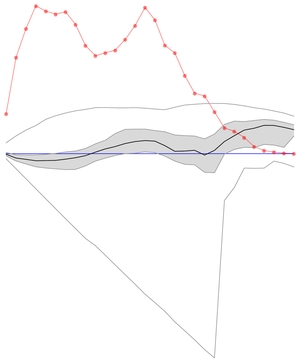} & \includegraphics[width=2.3cm]{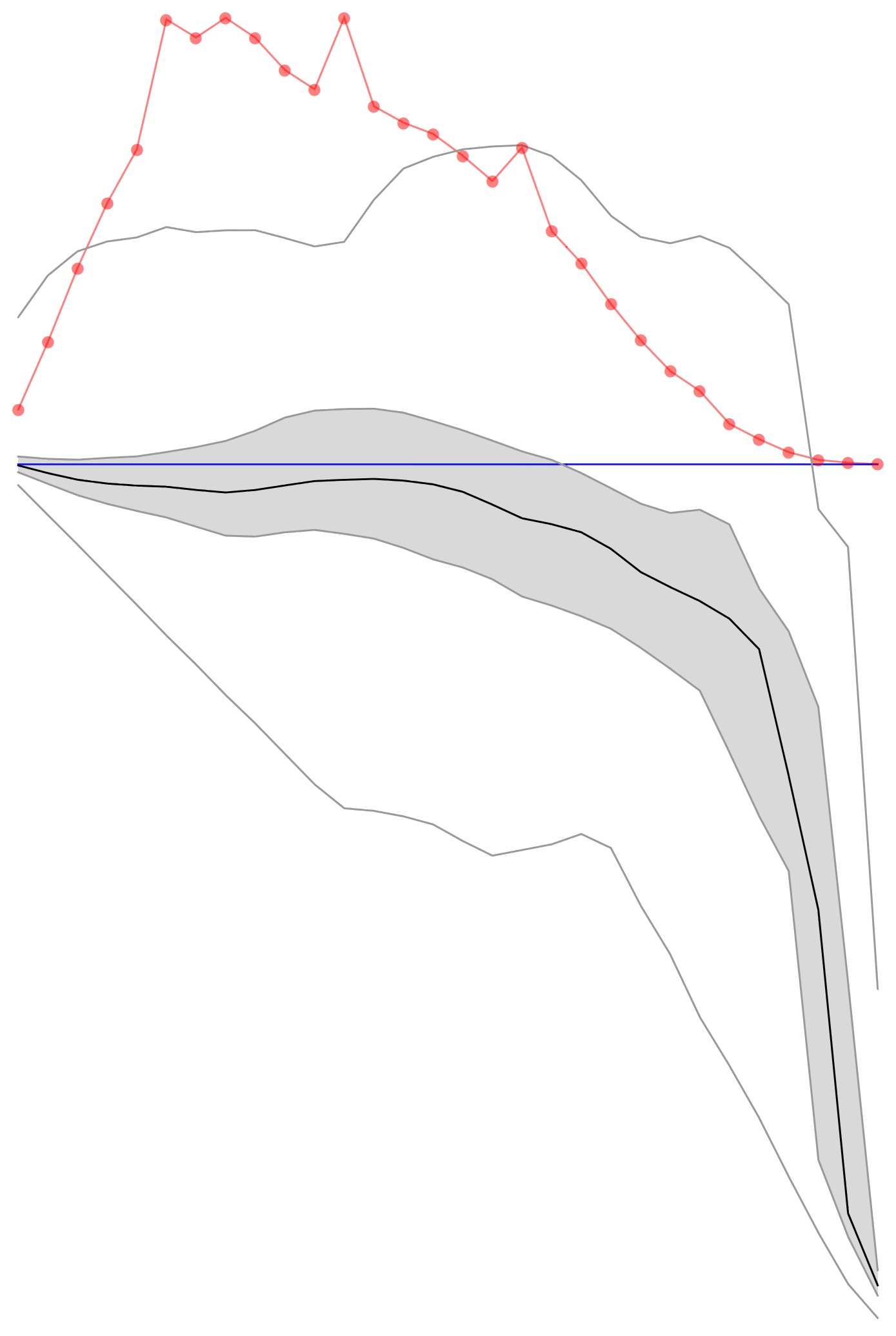}& \includegraphics[width=2.3cm]{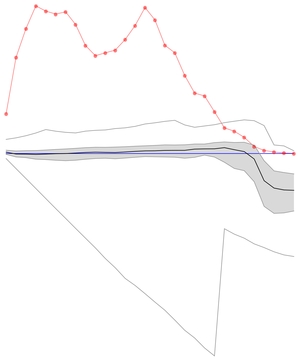}\\
\hline				   		 
 \raisebox{0.3in}[0pt]{\text{{\tt bcsstk31}}} & \includegraphics[width=2.3cm]{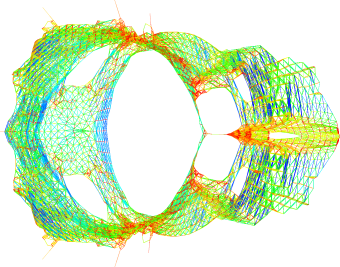} &\includegraphics[width=2.3cm]{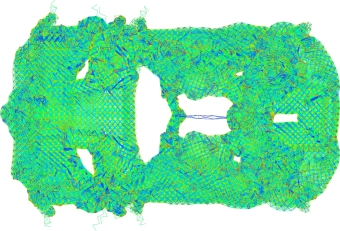}&\includegraphics[width=2.3cm]{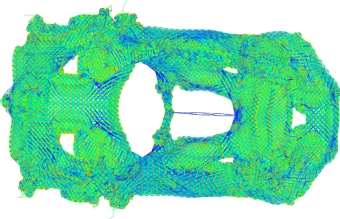}& \includegraphics[width=2.3cm]{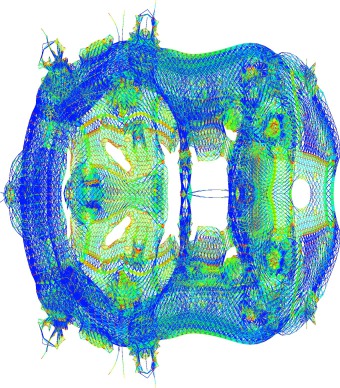}& -\\
\hline				   		 
 \raisebox{0.3in}[0pt]{\text{{\tt Luxembourg}}} & \includegraphics[width=2.3cm]{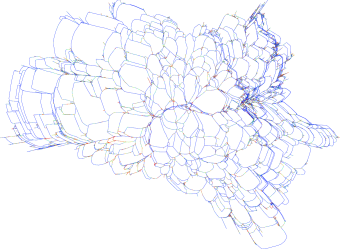} &\includegraphics[width=2.3cm]{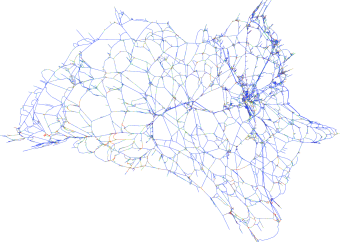}&\includegraphics[width=2.3cm]{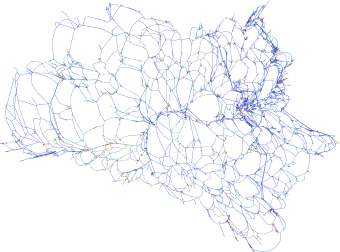}& \includegraphics[width=2.3cm]{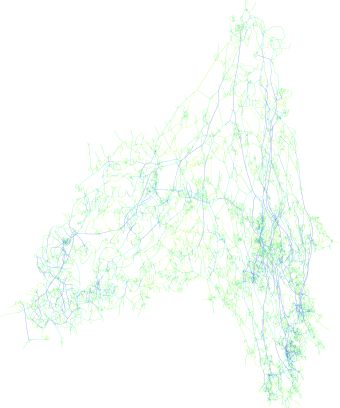} & -\\
\hline				   		 
\end{tabular}
\end{table}

In the graph renderings, we use a red-to-green-to-blue color
scale to encode edge lengths from short to long.
Edges shorter than half of the median edge length are red, 
edges longer than 1.5 times the median are blue, and other edges
are colored according to the scale.

With the error chart, we also include
a graph distance distribution curve (red), representing the number of 
vertex pairs in each graph distance bin. This distribution depends on the graph,
and is independent of the drawing. In making the error charts, the
layout is scaled to minimize the full stress model (\ref{spring_model}), with $w_{ij} = 1/d_{ij}^2$.

As an example, the error chart for PivotMDS on {\tt btree} (Table~\ref{btree_key},
column 1, bottom) shows that, on average, the median line is under the $x$-axis
for small graph distances. This means that the PivotMDS layout
under-represents the graph distance between vertex pairs that are a few
hops away. This is because it collapses branches of tree-like
structures. The leaves of such structures tend to be a few hops
away, but are now positioned very near to each other.  To some
extent the same under-representation of graph distance for vertex
pairs that are a few hops away is seen for PivotMDS and PivotMDS(1) on
other non-rigid graphs, including {\tt 1138\_bus}, {\tt btree}, {\tt lp\_ship041} and {\tt
USpowerGrid}.  Compared with PivotMDS and PivotMDS(1), the median line
for Maxent (column 3) does not undershoot the $x$-axes as much.

Comparing the COAST layouts with the others, we note that it appears to
track the $x$-axis more tightly and uniformly than the others, except for large lengths where,
in certain cases, it dives significantly.
In general, COAST has a more consistent bias for under-representation than the other layouts.
The others tend to under-represent short lengths and over-represent long lengths.
Visually, most of the COAST layouts are satisfactory, certainly avoiding the limitations
of PivotMDS. For example,
although it does not capture the symmetry of {\tt btree} as well as Maxent, it does a better job
of handling the details. 

\begin{comment}
The three largest graphs clearly differentiate COAST from the other non-FSM algorithms. Although it
does a good job separating the blades in the main and tail rotors, it appears to have a very difficult
time finding the relatively smooth grid surfaces. With {\tt bcsstk31} and {\tt luxembourg}, 
although we have no FSM layout for comparison, COAST is clearly the outlier. We hope that future work
can explain these anomalies.

As a side note, these error charts are helpful in understanding 
the characteristics of other
algorithms as well. For example,
...
\end{comment}

While visually comparing drawings made by different algorithms is
informative, and may give an overall impression of the characteristics
of each algorithm, such inspection is subjective. Ideally we would prefer
to rely on a quantitative measure of performance. However such a
measure is not easy to devise. For example, if we use sparse stress as
our measure, PivotMDS, which minimizes sparse stress, is likely to
come out best, despite its shortcomings. As a compromise, we propose to
measure full stress, as defined by (\ref{spring_model}), with
$w_{ij}=1/d_{ij}^2$. Bear in mind that this measure naturally favors
the full stress model. 

Table~\ref{stress_measure} gives the full stress
measure achieved by each algorithm, as well as the corresponding timings. Because it is expensive to
calculate all-pairs shortest paths, we restrict experimental measurement
to graphs with less than $10,000$ nodes. From the table we can see that, as
expected, FSM is the best, because it tries to optimize this measure.
We note that COAST is mostly competitive with the other non-FSM layouts. 
\begin{comment}
The exception is the  the {\tt commanche} example, where COAST is a factor of
two to three worse than the others, and about seven times worse than FSM.
This is probably related to what we remarked earlier about the respective 
drawings (\cite{gansner_hu_shankar_arxiv_2013}).
\end{comment}

As for timings, COAST, although a hybrid implementation, is comparable with Maxent, and
appears to work well on large graphs.
%Indeed, we note that the reason that COAST is not as fast for smaller graphs is that there is
%a fixed overhead in solving the optimization problem, which is dependent only on $k$.

   \begin{table*} 
    \caption{\small\sf Full stress measure ($\times 1000$) and CPU time (in seconds) for  PivotMDS, PivotMDS(1), Maxent, COAST and FSM. Smaller is better.
We limit the measurements to graphs with less than $10,000$ nodes and 10 hours of CPU time.
A ``-'' is used to denote these missing data points.
\label{stress_measure}}
     \begin{center}
     \begin{tabular}{|c||r|r||r|r||r|r||r|r||r|r|}
     \hline   
 \text{Graph} &   \multicolumn{2}{c||}{\text{PivotMDS}} & \multicolumn{2}{c||}{\text{PivotMDS(1)}} & \multicolumn{2}{c||}{\text{Maxent}} & \multicolumn{2}{c||}{\text{COAST}} & \multicolumn{2}{c|}{\text{FSM}} \\ 
\hline
 \text{{\tt gd}}          & 19 & 0.3  &   15 & 0.3 &   12 & 0.8 & 13 & 4.6 &   10 & 2.3 \\ 
\hline
 \text{{\tt btree}}       & 130 & 1.1  &  110 & 1.1  &   64 & 2.7  &    89 & 0.4 &  60 & 10.0 \\ 
\hline				   		 

\text{{\tt 1138\_bus}}   & 78 & 0.1  &   67 & 0.2 &   45 & 2.1 &    58 & 3.4 &  40 & 16.0 \\ 
\hline				   		 
 \text{{\tt qh882}}       & 147 & 0.1 &  120 & 0.3 &  103 & 2.2  &   184 & 2.7 &  84 & 39.0 \\ 
\hline				   		 
 \text{{\tt lp\_ship04l}} & 667 & 0.1 &  769  & 0.1 &  363 & 2.2  &   368 & 5.0 & 251 & 58.0 \\ 
\hline				   		 
 \text{{\tt USpowerGrid}} & 1124 & 0.1 &  932  & 0.9 & 1018 & 6.5 &   1073 & 5.3 & 702 & 272.0 \\ 
\hline	
 \text{{\tt commanche}}   & 2305 & 0.2 & 1547  & 0.9 & 1545  & 9.0 &  2853 & 8.8 & 654 & 1025.0 \\
\hline				   		 
\text{{\tt bcsstk31}}     & - & 2.4  &  - & 21.6  & - & 102.0 & - &  226.7 & - &  -\\
\hline				   		 
 \text{{\tt luxembourg}}  & - & 2.4  & - & 630.0    & - & 209.0 &  - & 128.9 & - &  -\\
\hline	
\end{tabular}
     \end{center}
     \end{table*}

%GD 13443.358765
%btree 89457
%1138_bus 57689
%qh882 184135
%%lp_ship01  368031
%USpowergrid 1073141
%commanche 2853125

\begin{comment}
     \begin{table*} 
    \caption{\small\sf CPU time (in seconds) for PivotMDS, PivotMDS(1), Maxent, COAST and FSM. 
A limit of 10 hour CPU time is imposed and ``-'' is used to denote runs that could not finish within that time, or ran out of memory.
\label{cpu}}
     \begin{center}
     \begin{tabular}{|c|c|c|c|c|c|c|c|c|c|c|c|c|c|c|c|c|c||c|c|c|c|c|c|}
     \hline   
 \text{Graph} &\text{PivotMDS} & \text{PivotMDS(1)} & \text{Maxent} & \text{COAST} &\text{FSM}\\
\hline
 \text{{\tt gd}}          & 0.3  &   0.3  & 0.8 & 10.7  &    2.3\\
\hline
 \text{{\tt btree}}       & 1.1  &   1.1  & 2.7 &  16.4 &   10\\
\hline
 \text{{\tt 1138\_bus}}   & 0.1  &   0.19 & 2.1 &  8.8 &   16\\
\hline				   		 
 \text{{\tt qh882}}       & 0.1  &   0.3  & 2.2 &  35.6 &   39 \\
\hline				   	 
\text{{\tt lp\_ship04l}}  & 0.1  &   0.1  & 2.2 &  52.0 &   58 \\
\hline				   		 
 \text{{\tt USpowerGrid}} & 0.1  &   0.9  & 6.5 &  82.1 &  272 \\
\hline				   		 
 \text{{\tt commanche}}   & 0.2  &   0.9  & 9.0 &  115.0 & 1025 \\
\hline				   		 
\text{{\tt bcsstk31}}     & 2.4  &  21.6  & 102 &  675.9 &   -\\
\hline				   		 
 \text{{\tt luxembourg}}  & 2.4  & 630    & 209 &  621.8 &   -\\
\hline	
\end{tabular}
     \end{center}
     \end{table*}
\end{comment}

%Compare with PMDS(k). k = 0: just pmds, k=1: sparse stress. k = 2: 2-neighborhood.
%Compare with sfdp.
%k-neighborhood

\subsection{Measuring Precision of Neighborhood Preservation}

Sometimes, in embedding high dimensional data into a lower dimension, one is interested in preserving 
the neighborhood structure.
In such a situation, exact replication of distances between objects becomes a secondary concern.

For example, imagine a graph where each node is a movie. Based on some recommender algorithm,
an edge is added between two movies if the algorithm predicts that a user who likes one movie would also 
like the other, with the length of the edge defined as the distance (dissimilarity) between the two movies. 
The graph is sparse because only movies that 
are strongly similar are connected by an edge. 
For a visualization of this data to be helpful,
we need to embed this graph in 2D in such a way
that, for each node (movie), nodes in its neighborhood
in the layout are very likely to be similar to 
this node. This would allow the user to explore movies that
are more likely of interest to her by examining, in the visualization, the
neighborhoods of the movies she knew and liked.

Following Gansner et al.~\cite{Gansner_Hu_North_maxent_tvcg_2013}, we look at 
the {\em precision of neighborhood preservation}.
We are interested in answering the question: if we see vertices
nearby in the embedding, how many of these are actually also
neighbors in the graph space? We define the precision 
of neighborhood
preservation as follows.  For each vertex $i$, $K$ neighboring
vertices of $i$ in the layout are chosen. These $K$ vertices are then
checked to see if their graph distance is less than a threshold
$d(K)$, where $d(K)$ is the distance of the $K$-th closest vertex to
$i$ in the graph space.  The percentage of the $K$ vertices that are
within the threshold, averaged over all vertices $i$, is taken as the
precision.  Note that precision (the fraction of retrieved instances that are relevant)
is a well-known concept in information retrieval.  Chen and Buja~\cite{Chen_2009_LMDS}
use a similar concept called {\em LC meta-criteria}.

\begin{figure*}[htbp]
\centering
\begin{tabular}{cc}
\includegraphics[width=0.5\textwidth]{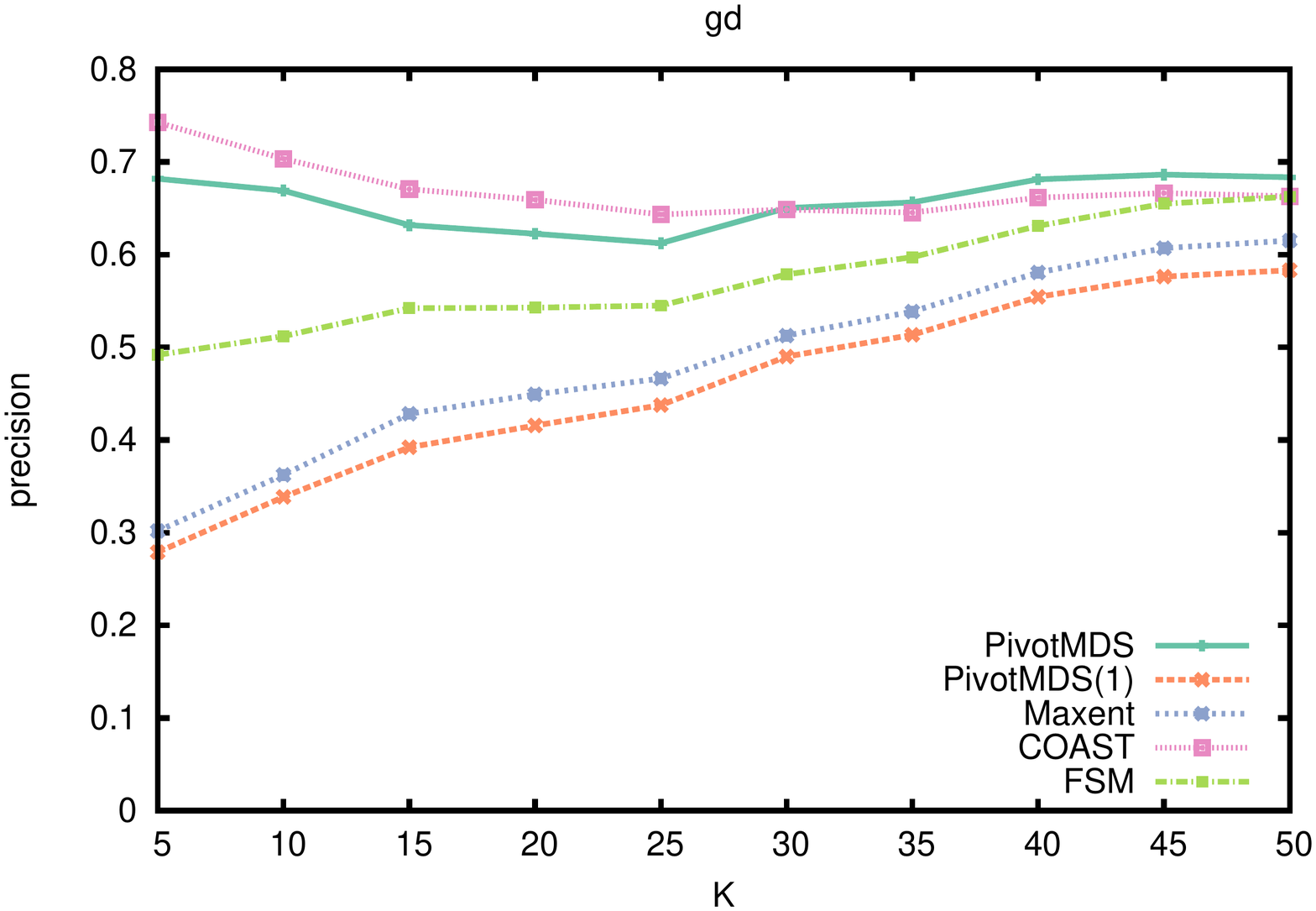}&
  \includegraphics[width=0.5\textwidth]{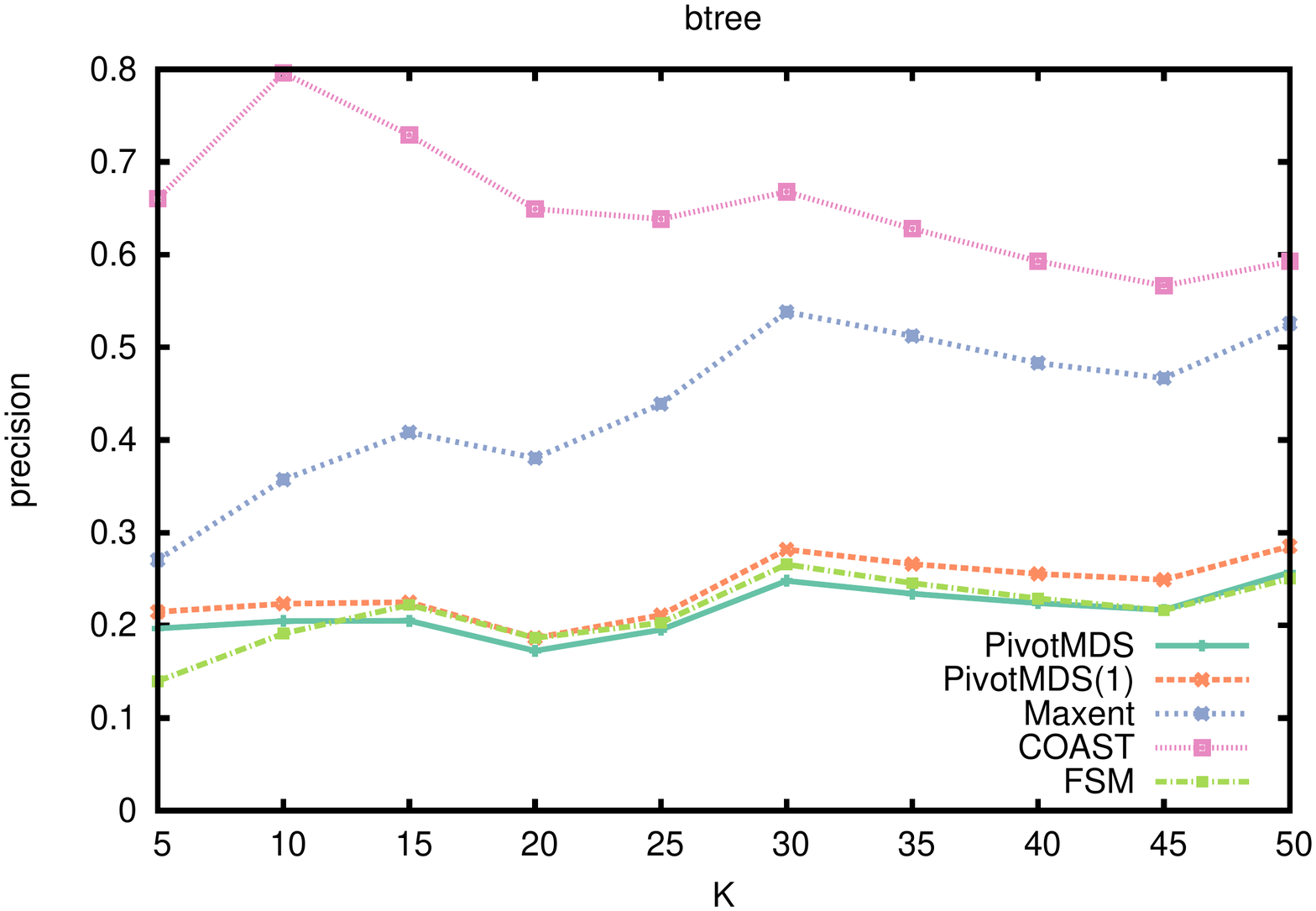}\\
\includegraphics[width=0.5\textwidth]{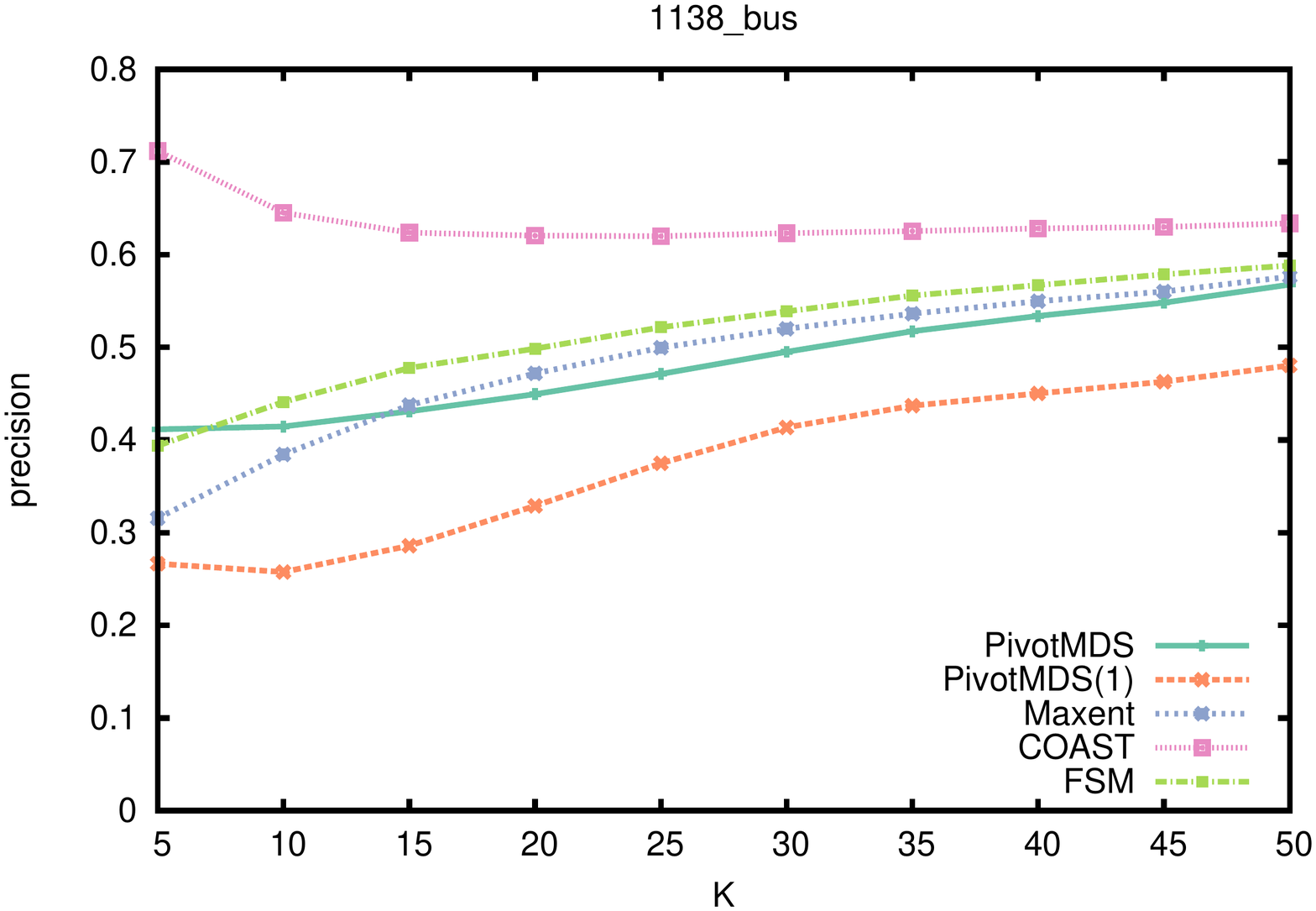}&
  \includegraphics[width=0.5\textwidth]{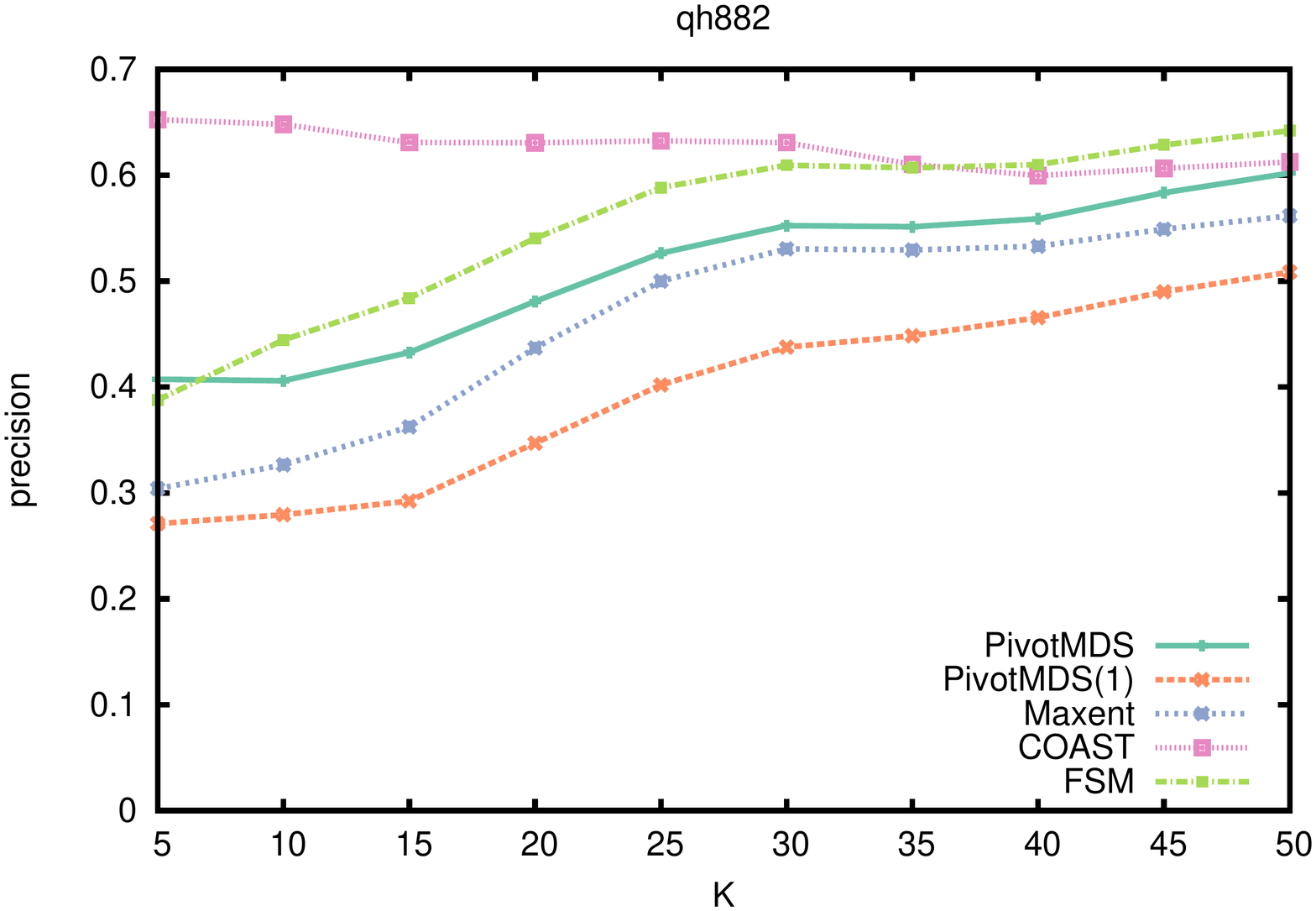}\\
\includegraphics[width=0.5\textwidth]{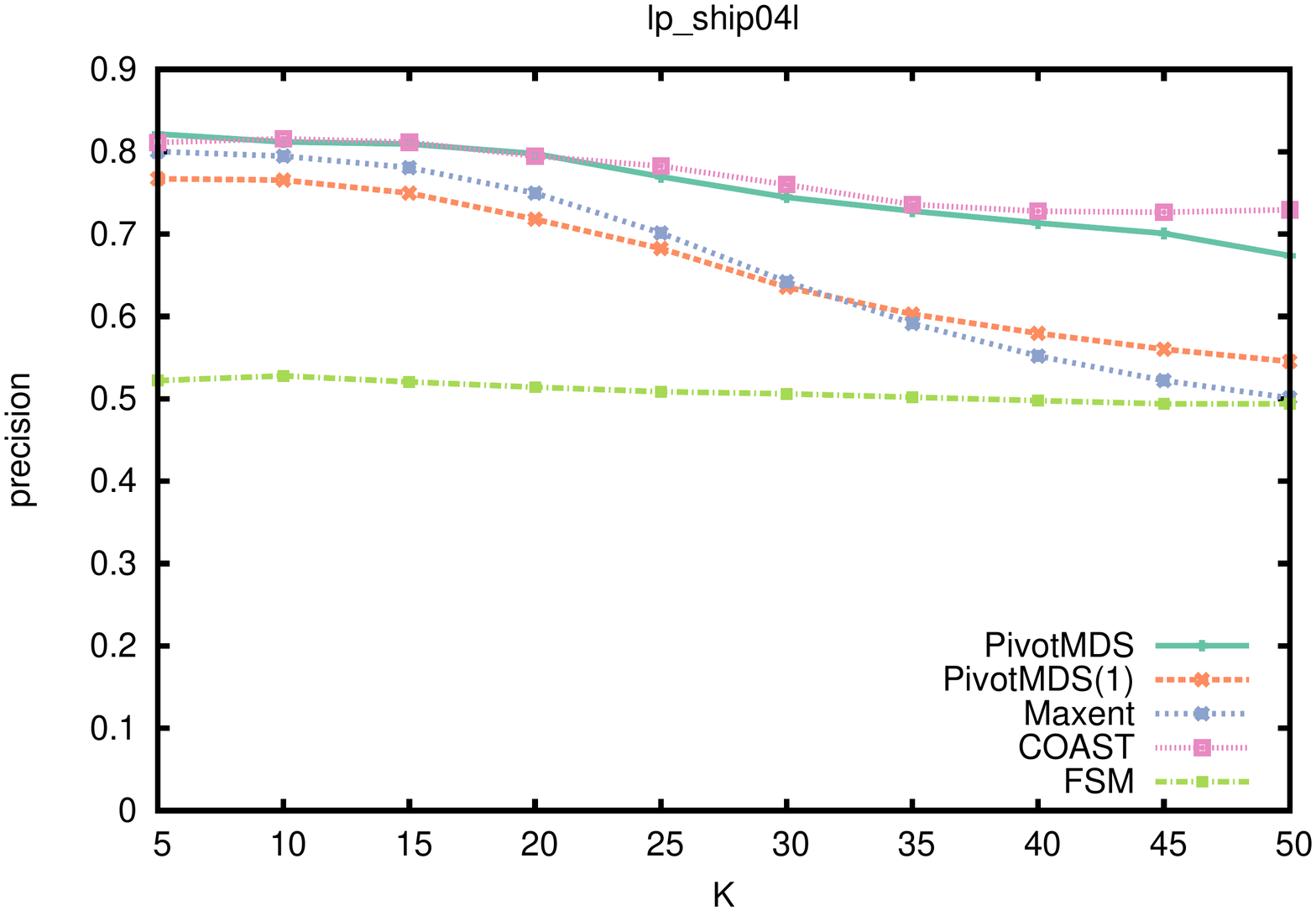}&
  \includegraphics[width=0.5\textwidth]{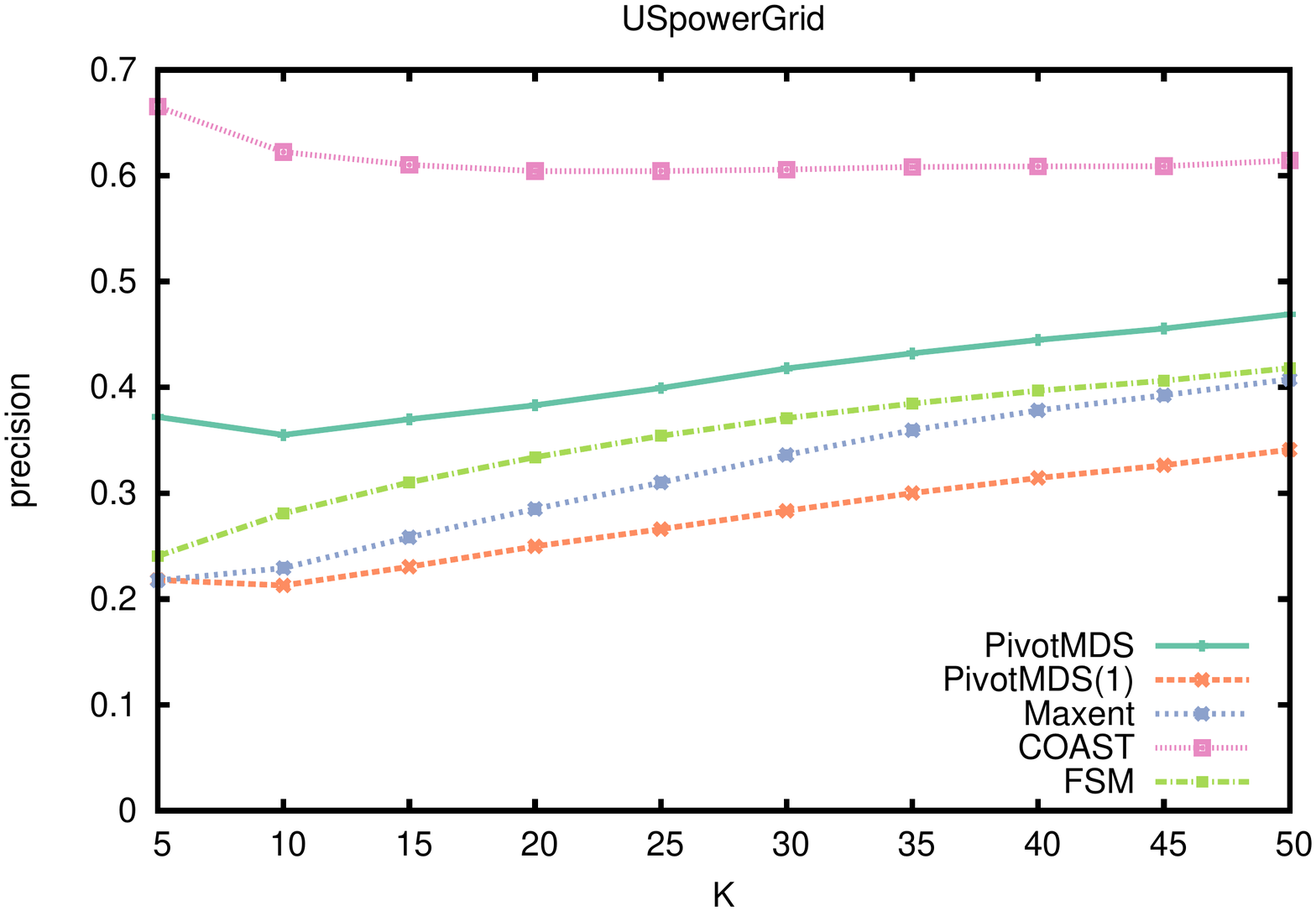}\\
\includegraphics[width=0.5\textwidth]{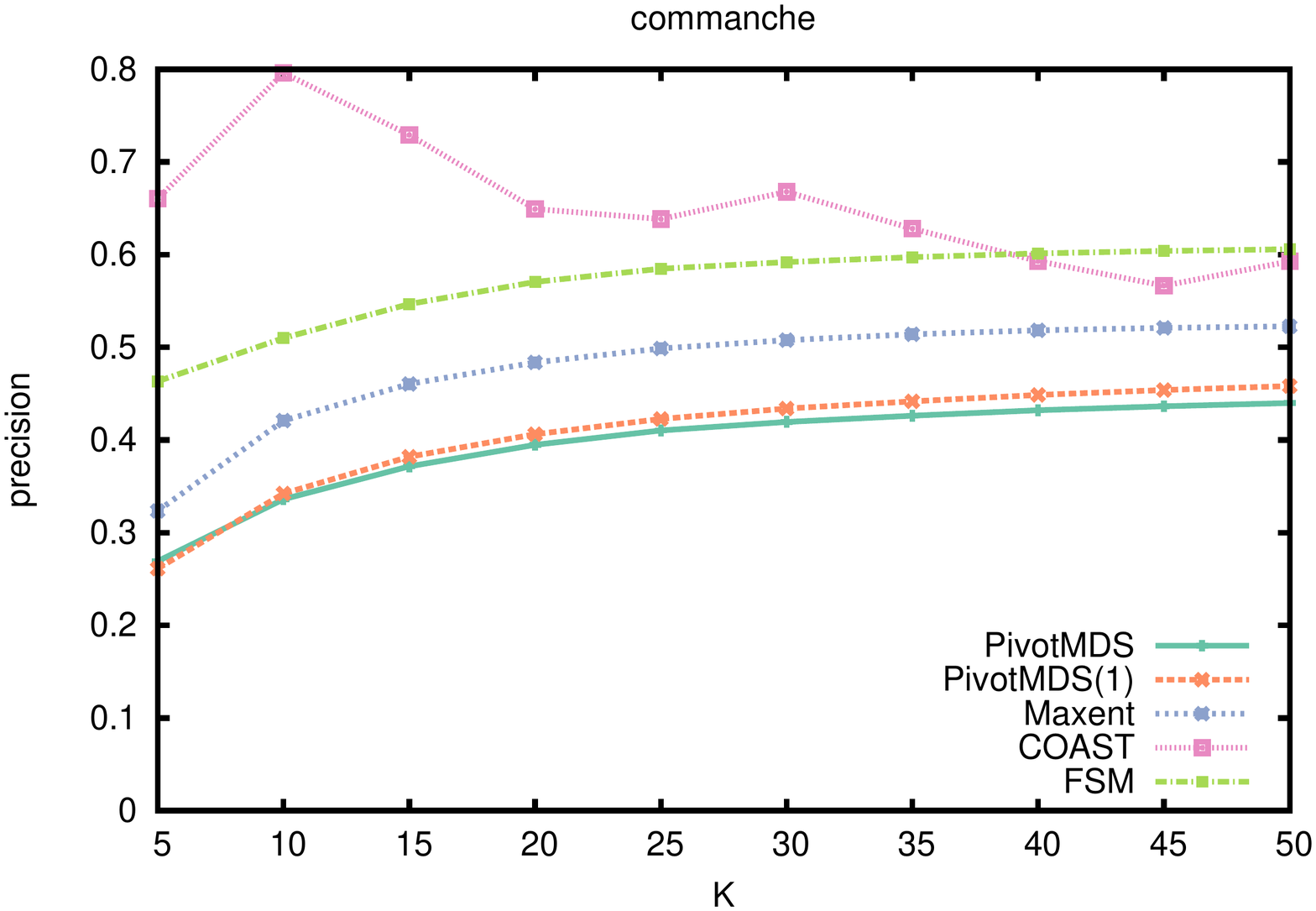}
\end{tabular}
\caption{Precision of neighborhood preservation of the algorithms, as a function of $K$. 
%For each vertex $i$, $K$-nearest neighbors
%of vertex $i$ in the layout
%is chosen. These $K$ vertices are then checked to see if their graph distance is less than a threshold $d(K)$, 
%where $d(K)$ is the graph distance of the $K$-th closest vertex to $i$ in the graph space.
%The percentage of the $K$ vertices that are within the threshold, averaged over all vertices $i$, is taken as the precision.
The higher the precision, the better.\label{fig_precision}}
\end{figure*}

Figure~\ref{fig_precision} gives the precision as a
function of $K$ for two representative graphs.
From the figure, it is seen that, in general, COAST has the highest,
or nearly the highest, precision.
PivotMDS(1) tends to have low precision. The precision
of other algorithms, including Maxent, tends to be between these two
extremes. 
%We do not have a good explanation for this, except to
%mention that in an earlier study~\cite{Gansner_Hu_North_maxent_tvcg_2013}, it was found that a force-directed algorithm 
%sfdp also has the highest precision.

%One outlier is {\tt lp\_ship04l}. For this case, PivotMDS(1)
%has high precision. In light of the drawing of this graph in
%Figure~\ref{lp}, this may be related to the fact that the drawing by PivotMDS(1), like that of
%PivotMDS, collapses the clusters.

%FSM has low precision for {\tt btree}. We believe this can be
%explained when looking at the drawings for {\tt btree} in
%Figure~\ref{drawings}. The FSM drawing utilizes space well, but
%each leaf $i$ of the tree tends to be very close to the leaves of
%other branches. These leaves have a high graph distance to $i$, making
%the precision lower. This is the result of FSM under-penalizing
%the under-prediction of large graph distances, as discussed before.

Overall, precision of neighborhood preservation
is a way  to look at one aspect of embedding 
not well-captured by the full stress objective function, but is important to 
applications such as recommendations.
COAST performs well in this respect.

\section{Conclusion and Future Work}\label{sec_conc}
In this paper, we described a new technique for graph
layout that attempts to satisfy edge length constraints. This technique uses
a modified two-part stress function, one part for the edge lengths, the
other to guide the relative placement of other node pairs. The stress is 
 quartic in the positions of the nodes, and 
can be transformed to a form that 
is suitable for solution using convex programming. The results produced are
good and the algorithm is scalable to large graphs.

Although the performance of the COAST algorithm is already competitive,
we rely on an {\it ad hoc} implementation using a combination of Python, Matlab and C code.
It would be very desirable to re-implement the algorithm entirely in C.
%, using the best available libraries. 
%with the most time consuming parts still in Matlab. 
%{\bf Is this still valid?}

Our technique follows 
the general strategy of doing length-sensitive drawings for large graphs
by reformulating the energy function, keeping the core length constraints, and then 
applying some appropriate mathematical machinery.
Variations of this technique have been successfully used by others~\cite{KhouryHKS_2012,Gansner_Hu_North_maxent_tvcg_2013}.
It would be interesting to explore additional adaptations of this approach.

\bibliographystyle{splncs03}
\bibliography{../ref}

%\appendix
%\clearpage
%\input{append}

\end{document}